\documentclass[11pt]{article}
\usepackage{amsmath}
\usepackage{scalerel}
\usepackage{amsthm,latexsym,amsxtra,graphicx,appendix,epstopdf,feynmf,setspace,fix-cm}
\usepackage[normalem]{ulem}
\usepackage{makeidx}
\usepackage{fullpage}
\usepackage{amssymb}
\usepackage{setspace}
\usepackage{bbm}
\usepackage{dsfont}
\usepackage{epsfig}
\usepackage[labelfont=bf,font=footnotesize,width=.94\textwidth]{caption}
\usepackage{cite}
\usepackage{multirow}
\usepackage{array,booktabs}
\usepackage{subfigure}
\usepackage{mathtools}
\usepackage{color}
\usepackage[makeroom]{cancel}
\usepackage{tikz}
\usepackage{pgfplots}
\usetikzlibrary{intersections, pgfplots.fillbetween}
\usepackage{tensor}
\usepackage{simplewick}
\usepackage{frcursive}
\usepackage{verbatim}
\usepackage{relsize}
\usepackage{cancel}
\pgfplotsset{compat=1.18}

\usepackage{pgfornament,multicol}

\usepackage{hyperref}

\usepackage[skip=0.2cm, indent=0pt]{parskip}

\hypersetup{
    colorlinks,%
    citecolor=blue,%
    filecolor=blue,%
    linkcolor=blue,%
    urlcolor=blue
}
\usepackage{cleveref}

\usepackage{float}
\usepackage{slashed}
\usepackage{tikz}
\usetikzlibrary{decorations.pathmorphing}

\renewcommand{\figurename}{Fig.}

\newcommand{\bi}{\begin{itemize}}
\newcommand{\ei}{\end{itemize}}
\newcommand{\bea}{\begin{eqnarray}}
\newcommand{\eea}{\end{eqnarray}}
\newcommand{\be}{\begin{equation}}
\newcommand{\ee}{\end{equation}}
\newcommand{\dd}{\text{d}}

\def\XXint#1#2#3{{\setbox0=\hbox{$#1{#2#3}{\int}$}
     \vcenter{\hbox{$#2#3$}}\kern-.5\wd0}}

\def\={\, = \,}

\makeatletter
\DeclareRobustCommand{\cev}[1]{%
  \mathpalette\do@cev{#1}%
}
\newcommand{\do@cev}[2]{%
  \fix@cev{#1}{+}%
  \reflectbox{$\m@th#1\vec{\reflectbox{$\fix@cev{#1}{-}\m@th#1#2\fix@cev{#1}{+}$}}$}%
  \fix@cev{#1}{-}%
}
\newcommand{\fix@cev}[2]{%
  \ifx#1\displaystyle
    \mkern#23mu
  \else
    \ifx#1\textstyle
      \mkern#23mu
    \else
      \ifx#1\scriptstyle
        \mkern#22mu
      \else
        \mkern#22mu
      \fi
    \fi
  \fi
}

\makeatother

\numberwithin{equation}{section}
\allowdisplaybreaks  

\begin{document}\spacing{1.2}

\pagenumbering{Alph}
\begin{titlepage}
  \thispagestyle{empty}

  \begin{flushright}
  \end{flushright}

\vskip3cm

\begin{center}  
{\Large\textbf{de Sitter at all loops: the story of the Schwinger model}}

\vskip1cm

\centerline{Dionysios Anninos,$^{\dagger\square}$\footnote{dionysios.anninos@kcl.ac.uk} 
Tarek Anous,$^\ddag$\footnote{t.anous@qmul.ac.uk} 
and Alan Rios Fukelman$^\dagger$\footnote{alan.rios\_fukelman@kcl.ac.uk}}
\vskip1cm

{\it{$^\dagger$Department of Mathematics, King's College London,
The Strand, London WC2R 2LS, United Kingdom}} 
\vskip 0.5cm
{\it{$^\square$Instituut voor Theoretische Fysica, KU Leuven, Celestijnenlaan 200D, B-3001 Leuven, Belgium}}
\vskip 0.5cm

{\it{$^\ddag$School of Mathematical Sciences, Queen Mary University of London,
Mile End Road, London E1 4NS, United Kingdom}}

\end{center}

\vskip1cm

\begin{abstract} 
We consider the two-dimensional Schwinger model of a massless charged fermion coupled to an Abelian gauge field on a fixed de Sitter background. The theory admits an exact solution, first examined by Jayewardena, and can be analyzed efficiently using Euclidean methods. We calculate fully non-perturbative, gauge-invariant correlation functions of the electric field as well as the fermion and analyze these correlators in the late-time limit.  We compare these results with the perturbative picture, for example by verifying that the one-loop contribution to the fermion two-point function, as predicted from the exact solution, matches the direct computation of the one-loop Feynman diagram.  We demonstrate many features endemic of quantum field theory in de Sitter space, including the appearance of late-time logarithms, their resummation to de Sitter invariant expressions, and Boltzmann suppressed non-perturbative phenomena, with surprising late-time features. 
\end{abstract}

\end{titlepage}
\pagenumbering{arabic}

\tableofcontents

\section{Introduction}
In this paper, we consider Schwinger's model \cite{Schwinger:1962tp,Schwinger:1963yda,Lowenstein:1971fc,Jackiw:1984zi,Coleman:1975pw,Coleman:1976uz,Roskies:1980jh} describing the two-dimensional quantum electrodynamics of a massless charged fermion $\Psi$ interacting with an Abelian gauge-field $A_\mu$. Due to the `unworldly' nature of two spacetime dimensions, the model, remarkably, is exactly soluble and exhibits several features of general interest from a four-dimensional perspective. This includes the spontaneous generation of a mass parameter for the gauge field, the presence of topological gauge-theoretic sectors, and the appearance of a chiral condensate. 

Although most studies of the Schwinger model place it on a Minkowski background, there exists a body of work reporting on its properties when placed on a general curved background  \cite{Gass:1982iu,Oki:1984tpr,Barcelos-Neto:1986oku,Jayewardena:1988td,Ferrari:1995uk}, where new phenomena of interest arise.
In what follows, we will analyze the Schwinger model placed on a two-dimensional de Sitter (dS$_2$) background, as well as its Euclidean counterpart, the $S^2$. Quantum field theory on the two-sphere captures the physics of the dS$_2$ invariant quantum state $|\text{E}\rangle$, i.e. the quantum field theoretic restriction to the Hartle-Hawking state. As such, it is invaluable to understand exactly soluble quantum field theories on the two-sphere. In particular, given $n$ points on $S^2$, the two-sphere $n$-point correlation function can be continued to a Lorentzian correlation in $|\text{E}\rangle$ at any spacetime point, provided one carefully treats the time-ordering prescription of the continuation. This continuation can often be hard to perform unless one has sufficiently complete results.

As was shown in the comprehensive treatment of \cite{Jayewardena:1988td}, one can indeed completely solve the Schwinger model on the round two-sphere. One is led to an effective Gaussian path integral that can be used to compute non-perturbative phenomena in exact form. As we shall see, the Euclidean effective action is partitioned by an integer $k \in \mathbb{Z}$ labeling distinct topological/instantonic configurations. The contribution of this article is to provide the generating functional of connected correlators of this model, and to point out that the results of \cite{Jayewardena:1988td} immediately provide us with \emph{all-loop} correlation functions in the Euclidean vacuum state of Lorentzian dS$_2$. We also directly compare these correlation functions to standard diagrammatic calculations in de Sitter space, which are often cumbersome to compute. We explicitly compute the late-time limits of our all-loop formulae and unveil their perturbative and non-perturbative structure. The model may be of relevance for recent efforts (see for example \cite{Bzowski:2013sza,Anninos:2014lwa,Baumann:2022jpr,Gorbenko:2019rza,DiPietro:2021sjt,Benincasa:2022gtd}) to understand quantum field theory in de Sitter space.

In section \ref{secmodel}, we introduce the Schwinger model on the two-sphere, while also presenting our choice of  geometric conventions for the $S^2$ and its analytic continuation dS$_2$. 
In section \ref{secEE}, we study the Euclidean vacuum two-point functions of the field-strength operator $F_{\mu\nu}$, at all loops, and explore its late-time and perturbative structure. In section \ref{secPsiPsi}, we extend this analysis to the  two-point function of the fermion operator $\Psi$. Finally, in the outlook section \ref{outlook} we discuss a composite scalar operator $:\bar{\Psi}\Psi:$, its properties, as well as some future directions. Our conventions are reported in \cref{ap:conventions} and we present details of certain important calculations in appendices \ref{app:PIandreg}, \ref{ap:integrateout}, and \ref{S2app}. 

\paragraph{Review versus new results: }This paper is a mix of review and new results. To distinguish between these two cases, let us indicate the review sections here, which are: section \ref{secmodel} (although our derivations differ from \cite{Jayewardena:1988td} in that we provide the full generating functional of correlators) as well as sections \ref{sec:exact2ptEM}, \ref{sec:nonvanishingtrace}, and \ref{sec:Calculation}. Readers wishing to see how we extract de Sitter physics from these results are invited to consult the complement of these sections, including appendices \ref{ap:integrateout} and \ref{S2app}.

\section{Schwinger model on a curved background}\label{secmodel}

The Schwinger model describes a two-component, massless, charged, Dirac spinor $\Psi$ interacting with a compact $U(1)$ gauge field $A_\mu$ in two spacetime dimensions. On a general Euclidean curved background $g_{\mu\nu}$ the Schwinger model is defined by the following action:
\begin{equation}\label{eq:Sschw}
    S_{\text{Schwinger}} =   \int  \textnormal{d}^2 x \,\sqrt{g}\left[\bar{\Psi} \gamma^\mu\left(\nabla_\mu  + i  A_\mu \right) \Psi+\frac{1}{4q^2} F^{\mu \nu} F_{\mu\nu}\right] ~ , 
\end{equation}
where $F_{\mu\nu} = \partial_\mu A_\nu - \partial_\nu A_\mu$ is the $U(1)$ field strength and $q$ is the gauge-coupling strength.  Note that $q$ has units of inverse length, implying that the theory is not scale invariant. At distances short compared to $1/q$, the theory is weakly coupled. In Euclidean signature, the two-component spinors $\bar{\Psi}$ and $\Psi$ are independent, but become related upon continuation to Lorentzian signature. Typically, unless otherwise stated, we will suppress spinor indices throughout. Our conventions are reported in \cref{ap:SpinorConventions}, including the definition of the spinor covariant derivative, as well as our conventions for the Clifford matrices $\gamma^\mu$.

\subsection{Symmetries and anomalies}
Being a gauge theory, this model has a set of invariances under local field reparametrizations. Under the following set of transformations: 
\begin{align}
&\Psi(x)\rightarrow h(x)\Psi(x)~, &\bar{\Psi}(x)\rightarrow \bar{\Psi}(x)h(x)^{-1}~,\nonumber\\ & A_\mu\rightarrow A_\mu+i h(x)^{-1}\partial_\mu h(x)^{}~, &h(x)=e^{i\alpha(x)}\in U(1)\label{eq:u1rot}~,
\end{align}
the action \eqref{eq:Sschw} remains unchanged. The classical theory, in fact, allows for a separate axial $U(1)$ \emph{global} symmetry, 
 which will be useful for later. Under the \emph{local} rotation
 \begin{equation}\label{eq:chiralrot}
\Psi(x)\rightarrow e^{i\beta(x)\gamma_*}\Psi(x)~, \qquad\qquad\bar{\Psi}(x)\rightarrow \bar{\Psi}(x)e^{i\beta(x)\gamma_*}~,
\end{equation}
the classical action transforms as
\begin{align}
S_{\text{Schwinger}}&\rightarrow S_{\text{Schwinger}}+i\int \dd^2x\sqrt{g}\,\left(\partial_\mu\beta\right)\left[\bar\Psi(x)\gamma^\mu\gamma_*\Psi(x)\right]\nonumber\\
&=S_{\text{Schwinger}}+\int \dd^2x\sqrt{g}\,\epsilon^{\mu\nu}\left(\partial_\mu\beta\right)\,\left[\bar\Psi(x)\gamma_\nu\Psi(x)\right]~,
\end{align}
where  $\gamma_*$ is the highest element of the Clifford algebra \eqref{gammast} and we have used \eqref{ap:2dcommute}. Note that for $\beta(x)=\text{const.}$, the action does not transform at all, as promised. 

It is well known that this classical symmetry does not survive in the quantum theory. This can be deduced by the fact  \cite{Fujikawa:1979ay, Roskies:1980jh,Jackiw:1984zi} that the  fermionic measure transforms under this transformation, and this transformation is ultimately responsible for the axial anomaly. Specifically, under \eqref{eq:chiralrot} we have \cite{Roskies:1980jh}: 
\begin{equation}
 D\bar\Psi D\Psi\rightarrow  D\bar\Psi D\Psi e^{+\frac{ i}{2\pi}\int \dd^2x\sqrt{g}\, \left( \beta(x)\epsilon^{\mu\nu}F_{\mu\nu}- i \beta(x) \nabla^2 \beta(x) \right)}~.
\end{equation}
Thus, taking account of the anomaly, under the rotation \eqref{eq:chiralrot}, the action transforms as 
\begin{equation}\label{eq:anomalychiralrot}
\delta S_{\text{Schwinger}} = \int \dd^2x\sqrt{g}\,\left\lbrace\epsilon^{\mu\nu}\partial_\mu\beta\,\left[\bar\Psi(x)\gamma_\nu\Psi(x)\right]-\frac{ i}{2\pi}\left( \beta(x)\epsilon^{\mu\nu}F_{\mu\nu} - i \beta(x) \nabla^2 \beta(x) \right)\right\rbrace~.
\end{equation}

Since we are interested in the Schwinger model on a maximally symmetric space with constant \emph{positive} curvature---in other words the two-sphere, we will give geometric preliminaries below. This case was studied extensively in \cite{Jayewardena:1988td}, although the connections to two-dimensional de Sitter space were left untouched. Indeed, the analytic continuation of the Euclidean $S^2$ to Lorentzian signature is none-other than the cosmological spacetime dS$_2$, which we will review below. This model's solvability on the $S^2$ makes it a remarkable playground for extracting exact non-perturbative lessons about interacting QFT in de Sitter. We will report on these lessons below.  

\subsection{Two-sphere and \texorpdfstring{dS$_2$}{ds2} geometry}

We now take a moment to introduce the background geometry,  which we take to be the round two-sphere of scalar curvature $R = +2/\ell^2$. This Euclidean spacetime can be defined via its embedding equation in $\mathbb{R}^3$, hence we will use symbols familiar from three-dimensional vector calculus, denoting vectors in $\mathbb{R}^3$ with an arrow, and dot products will always be taken with the flat Euclidean norm. For example, in terms of $\vec{\mathbf{r}}\in\mathbb{R}^3$, the embedding equation for the $S^2$ reads: 
\begin{equation}
\vec{\mathbf{r}}\cdot{\vec{\mathbf{r}}}=\ell^2~.
\end{equation} 
There are two coordinatizations of the sphere that we will typically use. One is the standard `round-'parametrization:
\begin{equation}
\vec{\mathbf{r}}=\ell\left(\cos\varphi\cos\vartheta,\sin\varphi\cos\vartheta,-\sin\vartheta\right)~, \quad\quad \vartheta \in (-\pi/2,\pi/2)~. 
\end{equation}
And the corresponding metric can be expressed as: 
\begin{equation} \label{eq:embmetric}
g_{\mu\nu}=\left(\partial_\mu\vec{\mathbf{r}}\right)\cdot\left(\partial_\nu\vec{\mathbf{r}}\right)~,\qquad\qquad x^\mu=( \vartheta,\varphi)~,
\end{equation}
giving the standard metric on the $S^2$:
\begin{equation}\label{sphere}
\frac{\dd s^2}{\ell^2} = \dd\vartheta^2+\cos^2\vartheta \dd\varphi^2 ~, \quad\quad \vartheta \in (-\pi/2,\pi/2)~. 
\end{equation}
More commonly, we will express the two-sphere metric in stereographic coordinates, represented by the map from $\mathbb{R}^2\rightarrow S^2$, given by the following:
\begin{equation}
\vec{\mathbf{r}} = \frac{2\ell^2}{\ell^2 +\mathbf{x}\cdot \mathbf{x}} \left( x^1, x^2,\pm\frac{\ell^2-\mathbf{x}\cdot \mathbf{x}}{2\ell}\right)~,
\label{coordstereo}
\end{equation}
where $\mathbf{x} = (x^1,x^2) \in \mathbb{R}^2$ denotes a point on the two-sphere and the choice of $\pm$ in \eqref{coordstereo} determines whether the origin corresponds to the south or north pole. Whenever an object is denoted by a bold symbol, but without an arrow, we mean this to evoke a two-vector rather than a three-vector. Thus, in these coordinates, we have: 
\begin{equation}\label{spherestereo}
\dd s^2= \frac{4 \ell^4\dd\mathbf{x}\cdot \dd \mathbf{x}}{(\ell^2+ \mathbf{x}\cdot \mathbf{x})^2}\equiv \Omega(\mathbf{x})^2\dd\mathbf{x}\cdot \dd \mathbf{x}~,
\end{equation}
and thus we see that the $S^2$ is Weyl-flat with conformal factor $\Omega(\mathbf{x})=\frac{2 \ell^2}{(\ell^2+ \mathbf{x}\cdot \mathbf{x})}$. For either choice of sign in \eqref{coordstereo}, the equator of the $S^2$ lies along $\mathbf{x}\cdot\mathbf{x}=\ell^2$. Unless otherwise stated, we work with the $+$ choice in \eqref{coordstereo}.

The $SO(3)$ invariant distance between two points on the sphere will be used extensively throughout. It can be expressed simply in terms of the embedding vectors: 
\begin{equation}\label{invso3}
\cos\Theta_{xy}\equiv \frac{\vec{\mathbf{r}}_x\cdot\vec{\mathbf{r}}_y}{\ell^2}~.
\end{equation}
We will also rely heavily on a second, although related, measure of the invariant distance, which we shall use extensively: 
\begin{equation}
u^E_{xy}\equiv\frac{\left(\vec{\mathbf{r}}_x-\vec{\mathbf{r}}_y\right)\cdot\left(\vec{\mathbf{r}}_x-\vec{\mathbf{r}}_y\right)}{2\ell^2}=2\sin^2\frac{\Theta_{xy}}{2}~, \label{eq:udef}
\end{equation}
which is adapted to analytic continuation to dS$_2$ and obeys $u^E_{x=y}=0$. By a simple identity, we note that 
\begin{equation}
\cos\Theta_{xy}=1-u^E_{xy}~.
\end{equation}

In the coordinate system (\ref{sphere}) one has 
\begin{equation}\label{SO3d}
u^E_{xy}=1-\left( \cos\left( \varphi_x - \varphi_y \right)\cos \vartheta_{{x}} \cos\vartheta_y + \sin\vartheta_x \sin\vartheta_y\right)~,
\end{equation}
whilst in stereographic coordinates one has
\begin{equation}
    u^E_{xy}=\frac{2\ell^2(\mathbf{x}-\mathbf{y})\cdot (\mathbf{x}-\mathbf{y})}{(\ell^2+\mathbf{x}\cdot\mathbf{x})(\ell^2+\mathbf{y}\cdot\mathbf{y})}~.
\end{equation}
We will sometimes write this in terms of complex coordinates: 
\begin{equation}\label{stereoCos}
u^E_{xy}=\frac{2\ell^2({z}_x -{z}_y)(\bar{z}_x - \bar{z}_y)}{(\ell^2+z_x\bar{z}_x)(\ell^2+z_y \bar{z}_y)}~, \quad\quad z_x \equiv x^1+i x^2~, \quad z_y \equiv y^1+i y^2~.
\end{equation}

\paragraph{Continuation to dS$_2$:} To obtain the metric on Lorentzian dS$_2$, we simply transform  $\vartheta \to i \tau$ in \eqref{sphere}, resulting in the global chart
\begin{equation}\label{global}
\frac{\dd s^2}{\ell^2} = -\dd\tau^2 + \cosh^2\tau \dd\varphi^2~, \quad\quad \tau \in\mathbb{R}~, \quad  \varphi \sim \varphi+2\pi~. 
\end{equation}
The de Sitter invariant distance in these coordinates is
\begin{equation}
u^L_{xy}=1+ \sinh\tau_x \sinh\tau_y- \cos\left( \varphi_x - \varphi_y \right)\cosh \tau_{{x}} \cosh\tau_y ~.\label{eq:desitterdistglob}
\end{equation}

The conformal global coordinates of dS$_2$ are given by
\begin{equation}\label{confglobal}
\frac{\dd s^2}{\ell^2} = \frac{-\dd T^2 +  \dd\varphi^2 }{\sin^2 T}~, \quad\quad T \in (-\pi,0)~, 
\end{equation}
which we obtain by starting from the metric \eqref{global} and performing the coordinate transformation $\cosh\tau=-\frac{1}{\sin T}$ and $\sinh\tau=-\cot T$ for $T\in (-\pi,0^-)$, as in \cite{Anninos:2023lin}. We note that in Lorentzian signature, and much unlike the two-sphere, the spacetime develops infinite asymptotia at $T= 0$ and $T=-\pi$ known as $\mathcal{I}^+$ and $\mathcal{I}^-$ respectively. In these coordinates, the de Sitter invariant distance takes the following form: 
\begin{equation}
u^L_{xy}=\frac{\cos(T_x-T_y)-\cos(\varphi_x-\varphi_y)}{\sin T_x\,\sin T_y}~.\label{eq:desitterdistconf}
\end{equation}

The two-point correlation functions computed in the Schwinger model on $S^2$ will often be functions of $u^E_{xy}$. To obtain the analytically continued dS$_2$ correlation functions, we will simply make the replacement $u^E_{xy}\rightarrow u_{xy}^L$.

\subsection{Topological sectors}

Before discussing the exact solution of the model \eqref{eq:Sschw}, let us first discuss the vacuum structure of $U(1)$ gauge theory in two-dimensions, in the absence of matter. We will start our discussion in Euclidean flat space $\mathbb{R}^2$ with metric $\dd s^2=\dd\mathbf{x}\cdot\dd\mathbf{x}$, after which we will move on to the case of $S^2$, whose metric differs only by a Weyl factor as in \eqref{spherestereo}. 

As should be familiar, gauge field configurations on $\mathbb{R}^2$ break up into sectors labeled by their \emph{winding number} (see chapter 7 of \cite{Coleman:1985rnk} for a classic introduction). The winding number is defined by the integral
\begin{equation}\label{eq:windingnumber}
-\frac{1}{4\pi}\int \dd^2x\sqrt{g}\, \epsilon^{\mu\nu} F_{\mu\nu}=k\in\mathbb{Z}~.
\end{equation}
Given its definition (see also \eqref{eq:levicivita} for our Levi-Civita conventions), it should hopefully be clear that this quantity is a topological-invariant, independent of the local geometry. And since the integrand is a total derivative, the  above integral can be evaluated by Stokes' theorem and computes an invariant quantity about the map from the $S^1$ at infinity into the compact gauge group $U(1)$---thus the winding number computes how many times the gauge field winds the $S^1$ at infinity.

It should be easy to verify that the following field configuration: 
\begin{equation}
A^k_\mu= k \,C^{(\lambda)}_\mu~, \qquad\qquad C^{(\lambda)}_\mu=\frac{1}{\lambda^2+\left(\mathbf{x}-\mathbf{y}\right)^2}\tilde{\epsilon}_{\mu\nu}\,\left(\mathbf{x}^\nu-\mathbf{y}^\nu\right)\label{eq:windingconfig}
\end{equation}
has winding number $k$. The quantity $\lambda$ is called the \emph{instanton size}, and $\mathbf{y}$ is the \emph{center of the instanton} and, in principle, both must be integrated over. The reason they must be integrated over is because the field configurations $A^k_\mu$ are off-shell, meaning they do not satisfy the free equations of motion, in other words:
\begin{equation}
\partial_\mu F^{\mu\nu}\neq0~.
\end{equation}
We can see this by computing the on-shell action for these configurations: 
\begin{equation}
\frac{1}{4q^2}\int_{\mathbb{R}^2} \dd^2x\sqrt{g}\, F_{\mu\nu}F^{\mu\nu}=\frac{2 \pi  }{3}\frac{k^2}{ q^2\lambda^2 }
\end{equation}
and noting that the minimum-energy configuration would be an instanton of infinite size, while nothing constraints the instanton's position. This causes a few issues in the flat-space story which we will not broach, however the interested reader may look at \cite{Adam:1993fc} to see this discussed in the context of the flat-space Schwinger model. The lesson is simply that the instanton configurations on $\mathbb{R}^2$ are not classical solutions upon which we can build a well-defined perturbative expansion. 

Now we can pass minimally to the $S^2$. Since the definition of the winding number is independent of the metric, \eqref{eq:windingconfig} still describes configurations of winding number $k$ on $S^2$. However, what changes is that the size of the instanton gets stabilized and its position gets set to the origin. Computing the on-shell action:  
\begin{equation}
\frac{1}{4q^2}\int_{S^2} \dd^2x\sqrt{g}\, F_{\mu\nu}F^{\mu\nu}=\frac{\pi  }{6}\frac{k^2}{ q^2\ell^2 }\left(1+\frac{2\,\mathbf{y}\cdot\mathbf{y}}{\ell^2}+\frac{\left(\ell^2+\mathbf{y}\cdot\mathbf{y}\right)^2}{\ell^2\lambda^2}+\frac{\lambda^2}{\ell^2}\right)~,
\end{equation}
we find it is minimized at $\lambda=\ell$ and $\mathbf{y}=0$. Indeed, computing $\nabla_\mu F^{\mu\nu}$ we find: 
\begin{equation}\label{eq:instlevelkaction}
\nabla_\mu F^{\mu\nu}=0~, \quad\quad\quad \frac{1}{4q^2}\int_{S^2} \dd^2x\sqrt{g}\, F_{\mu\nu}F^{\mu\nu}=\frac{\pi  }{2}\frac{k^2}{ q^2\ell^2 }~, \qquad\qquad \text{if } \lambda=\ell\text{ and } \mathbf{y}=0~.
\end{equation}
Thus we see that, on the $S^2$, we have a well-defined perturbative picture in each winding-number sector of size $\lambda=\ell$. Since it will play a prominent role in what is to come, we define: 
\begin{equation}\label{eq:sizeoneinstanton}
C_\mu\equiv C_\mu^{(\lambda=\ell)}=\frac{1}{\ell^2+\mathbf{x}\cdot\mathbf{x}}\tilde{\epsilon}_{\mu\nu}\,\mathbf{x}^\nu=-\frac{1}{2}\epsilon_{\mu\nu}\partial^\nu\log\Omega(\mathbf{x})~.
\end{equation} 
For an embedding-space construction of $C_\mu$, we invite the reader to consult \cref{ap:monopole}. Note that this vector potential can be thought of as a unit-charged magnetic monopole placed at the origin of the interior of the two-sphere in the embedding space $\mathbb{R}^3$, the location of the Dirac string being dictated by our choice of origin in stereographic coordinates (see \eqref{coordstereo}).

\paragraph{Topological caveat:} The attentive reader will have noticed that $\mathbb{R}^2$ and $S^2$ are topologically distinct spaces, and in reality there is no `$S^1$ at infinity' to speak of on the $S^2$. So why is the winding number still relevant in this context? For the sake of pedagogy, we have so far steered clear of topological issues, but let us address them here, following \cite{Jayewardena:1988td}. The hairy ball theorem \cite{poincare1885courbes} states that it is impossible to write a globally-defined everywhere-smooth and non-vanishing vector field on the $S^2$, meaning we must instead work in patches (see, for example, chapter 12 of \cite{Green:2012pqa}). Now consider a smooth gauge-field configuration that takes values $A_\mu(\mathbf{x})$ in the northern hemisphere and $A'_\mu(\mathbf{x}')$ in the southern hemisphere (to be explicit, choose $x^\mu$ and $x'^\mu$ to be the coordinates \eqref{coordstereo} with $\pm$ choices respectively). Using Stokes' theorem, we can write the winding number as a line integral along the equator:
\begin{equation}
-\frac{1}{4\pi}\int d^2x\sqrt{g}\, \epsilon^{\mu\nu} F_{\mu\nu}=-\frac{1}{2\pi}\left(\oint_{\mathbf{x}\cdot\mathbf{x}=\ell^2}A_\mu(\mathbf{x}) \dd x^\mu-\oint_{\mathbf{x}'\cdot\mathbf{x}'=\ell^2}A_\mu'(\mathbf{x}') \dd x'^\mu\right)~,
\end{equation}
where both integrals are computed in the same orientation. Since we are working in patches, it must be the case that the gauge fields $A_\mu(\mathbf{x})$ and $A'_\mu(\mathbf{x}')$ are related to each other by a gauge transformation in any overlap regions.  Along the equator it must therefore be that: 
\begin{equation}
A'_\nu(\mathbf{x}')= \left(A_\mu(\mathbf{x})+i h(\mathbf{x})^{-1}\partial_\mu h(\mathbf{x})\right)\frac{\partial x^\mu}{\partial x'^\nu}~.
\end{equation}
Plugging this back into the expression for the winding number, we find: 
\begin{equation}
-\frac{1}{4\pi}\int d^2x\sqrt{g}\, \epsilon^{\mu\nu} F_{\mu\nu}=\frac{i}{2\pi}\oint_{\mathbf{x}\cdot\mathbf{x}=\ell^2}\dd x^\mu\, h(\mathbf{x})^{-1}\partial_\mu h(\mathbf{x})~.
\end{equation}
Since $h\in U(1)$, we find that the winding number is an integer which now measures the number of times the \emph{transition function} winds around as we traverse the great circle at the equator. Hence, in order to path-integrate over all possible globally-defined field configurations on the $S^2$, we need to sum over the distinct winding number sectors, the same as we do in QCD in four-dimensional flat space.

A further historical comment is in order. One of the reasons that led \cite{Jayewardena:1988td} to consider the Schwinger model on the Euclidean $S^2$ was so that the author could provide a rigorous path-integral derivation of the known results in flat space (by taking $\ell\rightarrow\infty$), particularly the existence of the chiral condensate \cite{Lowenstein:1971fc}. It was already known \cite{Nielsen:1976hs,Rothe:1978hx} that a naive path-integral treatement of the Schwinger model on $\mathbb{R}^2$, in the zero-instanton sector, failed to reproduce the known result of chiral-symmetry breaking, originally derived using operator methods. The authors of \cite{Nielsen:1976hs,Rothe:1978hx} concluded that non-trivial instanton configurations needed to be included in the path integral in order to reproduce the non-perturbative nature of the chiral condensate. This intuition was later confirmed by the rigorous analysis in \cite{Jayewardena:1988td} (see also \cite{Adam:1993fy,Adam:1993fc,Adam:1994by} for later treatments on the $\mathbb{R}^2$).

\subsection{Exact solution} 

We now have assembled all the ingredients necessary to discuss the exact solution of the model following \cite{Jayewardena:1988td, Adam:1993fc,Adam:1994by,adam1993thesis}. We first need to parameterize our gauge field configurations. We make the following choice:
\begin{equation}\label{Ak}
A^{(k)}_\mu(\mathbf{x}) = k C_\mu(\mathbf{x}) +\epsilon_{\mu\nu} \partial^\nu\Phi(\mathbf{x}) + {i} h(\mathbf{x})^{-1} \partial_\mu h(\mathbf{x})~.
\end{equation}
The first term represents the $k^{\rm th}$ topological sector background, whereas the middle term is a fluctuation satisfying the Lorenz gauge condition, and we take $\Phi(\mathbf{x})$ to have winding number zero. It is important to note that the path integral over the field $\Phi(\mathbf{x})$ does not include a zero-mode, as constant shifts of $\Phi$ do not change the field configuration. The last term represents the pure-gauge configurations, which drop out when computing any physical observable. In changing the parameterization of the general gauge field configuration $A^{(k)}_\mu(\mathbf{x})$ to $\{C_\mu(\mathbf{x}),\Phi(\mathbf{x}), h(\mathbf{x})\}$, the path integration measure over the $\{\Phi(\mathbf{x}), h(\mathbf{x})\}$ will be accompanied by a Jacobian factor $J_{\Phi,h}$ \cite{Giombi:2015haa} given by a functional determinant of the scalar Laplacian with a zero-mode removed. Generally this Jacobian drops out of physical observables but plays a role in the two-sphere path integral that we compute in appendix \ref{S2app}.

We are interested in computing the path integral with sources, which splits up into a sum over topological sectors:\footnote{For simplicity we omit the possibility of a $\theta$-angle, although there is no obstruction to including one.} 
\begin{equation}
Z[\eta, \bar{\eta},J_\mu]=\sum_{k=-\infty}^{\infty}Z_k[\eta,\bar{\eta},J_\mu]
\end{equation}
where
\begin{equation}\label{eq:fullzk}
Z_k[\eta,\bar{\eta},J^\mu]=\int \frac{D\bar{\Psi}D\Psi D\Phi {D}h}{\textnormal{vol}\mathcal{G}} J_{\Phi,h}\, e^{-\int  \textnormal{d}^2 x \,\sqrt{g}\left[\bar{\Psi} \gamma^\mu\left(\nabla_\mu  + i  A^{(k)}_\mu\right) \Psi+\frac{1}{4q^2} F^{\mu \nu} F_{\mu\nu}-J^\mu A^{(k)}_\mu-\bar\Psi\eta-\bar\eta\Psi\right]}
\end{equation}
and $J_{\Phi,h}$ is the Jacobian from the change of variables \eqref{Ak} and we have included a source for the gauge field $J^\mu$ as well as sources for the fermionic fields $\eta$ and $\bar{\eta}$~.

To simplify the analysis of this theory, we first perform a $U(1)$ rotation on the fermions \eqref{eq:u1rot} to remove any reference to $h$ in the Dirac action, followed by a chiral rotation (\ref{eq:chiralrot}) with $\beta(\mathbf{x}) = + i \Phi(\mathbf{x})$. This judicious choice removes the interaction between the fermion and the transverse part of the gauge field. As we will now show, the cost of performing this rotation is the spontaneous generation of a mass term for the transverse gauge field $\Phi(\mathbf{x})$, whose origin stems from the chiral anomaly \eqref{eq:anomalychiralrot}. A straightforward calculation (including integration by parts) yields the following kinetic term for $\Phi(\mathbf{x})$:
\begin{equation}\label{eq:Sphinotsimplified}
    S_\Phi =  \frac{\pi k^2}{2 q^2 \ell^2} + \frac{k}{2 q^2\ell^2} \int \dd^2 x \sqrt{g}\left( \nabla^2 - \frac{q^2}{\pi} \right) \Phi + \frac{1}{2 q^2} \int \dd^2 x \sqrt{g} \Phi \nabla^2\left( \nabla^2 - \frac{q^2}{\pi}  \right) \Phi  ~, 
\end{equation}
where $\nabla^2\equiv \frac{1}{\sqrt{g}}\partial_\mu \left(\sqrt{g}g^{\mu\nu}\partial_\nu\right)$ is the usual scalar Laplacian.

Let us now unpack this expression. The first term is the on-shell action for the $k$-instanton configuration, computed in \eqref{eq:instlevelkaction}. The last term is the kinetic term for $\Phi$, which we note is higher-derivative in nature and clearly does not produce a linear dispersion relation. Rather, we immediately identify a mass parameter $m^2 \ell^2 = \frac{q^2 \ell^2}{\pi}$ originating from the chiral anomaly. Finally, the middle term is a cross-term between the $k$-instanton background and the fluctuation $\Phi$. This term vanishes for two reasons. The first piece of the cross-term is a total derivative and therefore contributes a surface term, which vanishes because the $S^2$ has no boundary. The second piece of the cross-term is proportional to the zero-mode of $\Phi$, which, as we already mentioned, is absent in this parametrization. Thus, we simply have: 
\begin{equation}\label{eq:Sphi}
    S_\Phi =  \frac{\pi k^2}{2 q^2 \ell^2} + \frac{1}{2 q^2} \int \dd^2 x \sqrt{g} \Phi \nabla^2\left( \nabla^2 - \frac{q^2}{\pi}  \right) \Phi  ~. 
\end{equation}

Now let us analyze the fermion kinetic term: 
\begin{equation}\label{eq:Sfer}
    S_\Psi = \int \textnormal{d}^2 x \sqrt{g} \bar{\Psi} \gamma^\mu \left( \nabla_\mu + i k C_\mu  \right) \Psi \, . 
\end{equation}
Recall that the Dirac action obeys a simple transformation rule upon Weyl-rescalings, where the fermions have Weyl-weight 1/2. We therefore exploit the stereographic coordinate system (\ref{coordstereo}), to write:
\begin{equation}
    S_\Psi = \int \dd^2 x\sqrt{g}\, \bar{\Psi} \slashed{\nabla}_k \Psi ~, 
\end{equation}
with 
\begin{equation}\label{eq:nablaslashedk}
    \slashed{\nabla}_k\equiv\Omega^{-2}\begin{pmatrix}\Omega^{\frac{1-k}{2}} & 0 \\ 0 & \Omega^{\frac{1+k}{2}}\end{pmatrix}\begin{pmatrix}0 &\partial_{x^1}-i\partial_{x^2} \\ \partial_{x^1}+i\partial_{x^2} &0\end{pmatrix}\,\begin{pmatrix}\Omega^{\frac{1-k}{2}} & 0 \\ 0 & \Omega^{\frac{1+k}{2}}\end{pmatrix}
\end{equation}
where we have absorbed the effect of the spin connection and background $k$-instanton gauge field into a similarity transformation of the flat-space Dirac operator. 

Given the Gaussian nature of the theory, we are now almost ready to integrate out the fermions entirely. However, we must note that when $k\neq 0$, the above Dirac operator $\slashed{\nabla}_k$ has $|k|$ zero-modes, which must be treated separately. The fact that there are only $|k|$ such zero-modes stems from the requirement that they be normalized according to: 
\begin{equation}
    \int_{S^2}\dd^2x\sqrt{g} \bar\Psi \Psi =\ell~.
\end{equation}
Denoting the zero-modes by $\chi_s$, for $s=1,\dots,|k|$, we have
\begin{align}
    \chi_s=\Omega^{\frac{|k|-1}{2}}\left(\frac{\bar{z}}{\ell}\right)^{s-1}\sqrt{\frac{|k|}{2^{1+|k|}\pi\ell}\binom{|k|-1}{s-1}}&\begin{pmatrix}0 \\ 1\end{pmatrix}~, &k<0~,\nonumber\\ 
    \chi_s=\Omega^{\frac{k-1}{2}}\left(\frac{z}{\ell}\right)^{s-1}\sqrt{\frac{k}{2^{1+k}\pi\ell}\binom{k-1}{s-1}}&\begin{pmatrix}1 \\ 0\end{pmatrix}~,  &k>0~,\label{eq:zeromodes}
\end{align}
where we have used the complex coordinates introduced in \eqref{stereoCos}.
We are finally ready to integrate out the fermion $\Psi$. In our goal of computing \eqref{eq:fullzk}, we must calculate
\begin{equation}
   Z^\Psi_k\equiv\int D \bar{\Psi} D \Psi e^{-\int  \textnormal{d}^2 x \,\sqrt{g}\left[\bar{\Psi} \slashed{\nabla}_k \Psi-\bar\Psi e^{-\Phi\gamma_*}h^{-1}\eta-\bar\eta h e^{-\Phi\gamma_*}\Psi\right]}~,
\end{equation}
where our sequence of rotations on the fermions $\Psi$ and $\bar{\Psi}$ have now appeared in the source term. The interested reader may consult \cref{ap:integrateout} for the details, but the answer is:
\begin{equation}
 Z^\Psi_{k,\tilde{\epsilon}}= \frac{\mathcal{N}^\Psi_{\tilde{\epsilon}}\,e^{\frac{k^2}{2}}}{ |k|! \prod_{l=1}^{|k|} l^{2l-|k|-1}}\left[ \prod_{j=1}^{|k|}\left(\hat{\overline{\eta}},\chi_j\right)\left(\bar{\chi}_j,\hat{\eta}\right)\right]e^{ \int \dd^2 z\, \dd^2 w \sqrt{g_z} \sqrt{g_w} \,\hat{\overline{\eta}}(\mathbf{z}) S_k(\mathbf{z},\mathbf{w}) \hat{\eta}(\mathbf{w})}\label{eq:fermitopological}
\end{equation}
where we have used the compact notation
\begin{equation}
\left(f,g\right)\equiv \int\dd^2 x\sqrt{g}\,  f(\mathbf{x}) g(\mathbf{x})   \, .
\end{equation}
We label the quantity in \eqref{eq:fermitopological} with a subscript $\tilde{\epsilon}$ to indicate that this is a scheme-dependent UV-regulated quantity. The overall normalization $\mathcal{N}_{\tilde{\epsilon}}^\Psi$, for example, depends on our specific regularization scheme and is given in \eqref{eq:normalizationfermdet}. This quantity will drop out in physical observables, however the $k$-dependence of the prefactor across the topological sectors is important. 

The quantity $S_k(\mathbf{x},\mathbf{y})$ appearing above is the Green function of the operator (\ref{eq:nablaslashedk}) defined in (\ref{SKResult}), which we present here in stereographic coordinates: 
\begin{equation} \label{eq:Skmaintext}
S_k(\mathbf{x},\mathbf{y})  = \frac{1}{2\pi}  \begin{pmatrix}\Omega(\mathbf{x})^{-\frac{1-k}{2}} & 0 \\ 0 & \Omega(\mathbf{x})^{-\frac{1+k}{2}}\end{pmatrix}\frac{\boldsymbol{\sigma}\cdot (\mathbf{x}-\mathbf{y})}{|\mathbf{x}-\mathbf{y}|^2}  \begin{pmatrix}\Omega(\mathbf{y})^{-\frac{1-k}{2}} & 0 \\ 0 & \Omega(\mathbf{y})^{-\frac{1+k}{2}}\end{pmatrix}~.
\end{equation}
Finally we have introduced the following rotated sources in the above expression:  
\begin{equation}
    \hat{\overline{\eta}}(\mathbf{x}) = \bar{\eta}(\mathbf{x}) h(\mathbf{x}) e^{-\Phi(\mathbf{x}) \gamma_*} \, , \qquad \hat{\eta}(\mathbf{x}) = e^{-\Phi(\mathbf{x})\gamma_*} h^{-1}(\mathbf{x}) \eta(\mathbf{x}) \, ~.
\end{equation}
The exact generating function of connected correlators of the theory is then given by 
{
\begin{multline} \label{exactSpf}
Z[\eta, \bar{\eta},J_\mu]=\mathcal{N}^\Psi_{\tilde{\epsilon}}\sum_{k=-\infty}^{\infty} \frac{\,e^{\frac{k^2}{2}\left(1- \frac{\pi }{q^2 \ell^2}\right)}}{ |k|! \prod_{l=1}^{|k|} l^{2l-|k|-1}}\int \frac{D \Phi D h}{\textnormal{vol}\mathcal{G}} J_{\Phi,h}\left[ \prod_{j=1}^{|k|}\left(\hat{\overline{\eta}},\chi_j\right)\left(\bar{\chi}_j,\hat{\eta}\right)\right]\\ \times e^{\int \dd^2 z\, \dd^2 w \sqrt{g_z} \sqrt{g_w} \,\hat{\overline{\eta}}(\mathbf{z}) S_k(\mathbf{z},\mathbf{w}) \hat{\eta}(\mathbf{w}) - \frac{1}{2 q^2} \int \dd^2 x \sqrt{g} \Phi \nabla^2\left( \nabla^2 - \frac{q^2}{\pi}  \right) \Phi  +\int \dd^2x\sqrt{g} J^\mu A_\mu^{(k)}} ~.
\end{multline}
It is truly remarkable that such an expression can be derived for a gapped interacting field theory on Euclidean de Sitter space. While it may not yet be obvious, in the next two sections we will demonstrate that this expression can be used to compute \emph{exact, fully non-pertubative} correlation functions of an arbitrary number of fermions and gauge bosons.}

We are now ready to exploit the exact solution (\ref{exactSpf}) to compute a variety of gauge-invariant observables in the theory. The simplest such observable is the zero-point function itself, whose computation we delineate in appendix \ref{S2app}. The result is an explicit expression \eqref{eq:Ztotal} that can be systematically expanded in a perturbative, small $q^2\ell^2$ expansion. In the following, we focus instead on the simplest correlation functions of the theory, namely the two-point function of the electric field and fermion operators.

\section{Electric field two-point function}\label{secEE}

In this section, we will consider the two-point function of the electric field operator $F_{12}(\mathbf{x})$. Using the parametrization (\ref{Ak}), the field strength is:
\begin{align}
    F_{\mu \nu}^{(k)}(\mathbf{x}) &= k \left(\partial_\mu C_{\nu }-\partial_\nu C_{\mu} \right) + \left[\partial_{\mu} \left(\epsilon_{\nu \rho}\partial^\rho \Phi(\mathbf{x})\right)-\partial_{\nu} \left(\epsilon_{\mu \rho}\partial^\rho \Phi(\mathbf{x})\right)\right]  \, ,\nonumber\\
    &=-\epsilon_{\mu\nu}\nabla^2_{\mathbf{x}}\left(\Phi(\mathbf{x})-\frac{k}{2}\log\Omega\right)~,
\end{align}
and note that any reference to the pure-gauge configuration $h(\mathbf{x})$ has dropped out, as expected.
If our aim is to compute bosonic correlation functions---that is, correlation functions with only bosonic operators inserted---then, as previously mentioned, the non-trivial topological sectors with $k \neq 0$ do not contribute once we set the sources $\eta$ and $\bar\eta$ to zero (see equation \eqref{eq:fermitopological}). In this section we will compute the two point functions of $F_{\mu \nu}(\mathbf{x})$ and thus we can readily set $k=0$. Since we are in two dimensions, the field strength only has one independent degree of freedom
\begin{equation}
    F_{12}(\mathbf{x}) = - \sqrt{g_x}\, \nabla^2_{\mathbf{x}} \Phi(\mathbf{x}) \, , 
\end{equation}
with $\Phi(\mathbf{x})$ the transverse component of the gauge field $A_\mu(\mathbf{x})$ in Lorenz gauge.

\subsection{Exact two-point function}\label{sec:exact2ptEM}

Given the Gaussian nature of the action (\ref{exactSpf}), it is straightforward to first compute the two-point function of $\Phi$ in the Euclidean state $|\text{E} \rangle$. We find
\begin{equation}
 G_\Phi(\mathbf{x},\mathbf{y}) \equiv  \langle \text{E}| \Phi(\mathbf{x})\Phi(\mathbf{y}) | \text{E} \rangle = q^2 \ell^2 \sum_{L=1}^\infty \sum_{M=-L}^L \frac{Y_{LM}(\mathbf{x})Y_{LM}(\mathbf{y})}{L (L+1) \left(L(L+1)+\frac{q^2\ell^2}{\pi} \right)} ~.
\end{equation}
To arrive at this expression, we have used standard spherical harmonics techniques that are reviewed in the appendix \ref{ap:sphH}. The sum does not include the $L=0$ term since $\Phi(\mathbf{x})$ has no zero-mode. Amazingly, we can calculate the above sum exactly by noticing that
\begin{equation}\label{eq:sumdiff}
\frac{q^2 \ell^2  }{L (L+1) \left(L(L+1)+\frac{q^2\ell^2}{\pi}\right)}=\pi\left(\frac{1}{L(L+1)}-\frac{1  }{L(L+1)+\frac{q^2\ell^2}{\pi}}\right)~,
\end{equation}
giving:
\begin{equation}
 G_\Phi(\mathbf{x},\mathbf{y}) =-\frac{1}{4}\left(1-\frac{\pi}{q^2\ell^2}\right)-\frac{1}{4}\left[\log \frac{u_{xy}^E}{2}+{\Gamma(\Delta) \Gamma(1-\Delta)}\, {_2}F_1\left(\Delta,1-\Delta,1,1-\frac{u_{xy}^E}{2}\right)\right]~,
\label{GF}
\end{equation}
where we remind the reader that $u_{xy}^E$ is the invariant geodesic distance on the $S^2$ defined in \eqref{eq:udef}. As is customary in the de Sitter quantum field theory literature, we have also defined the `conformal weight' $\Delta$ and corresponding quadratic Casimir of the dS$_2$ symmetry group, which is related to the coupling strength $q^2$ as follows:
\begin{equation}
\Delta \equiv \frac{1}{2} + i \sqrt{\frac{q^2\ell^2}{\pi} - \frac{1}{4}}~, \quad\quad\quad \Delta(1-\Delta) = \frac{q^2 \ell^2}{\pi}~.
\end{equation}

The Green's function $G_\Phi$ satisfies the following differential equation:
\begin{equation}\label{eq:gphidiffeq}
- \frac{1}{q^2} \nabla^2 \left(-\nabla^2 + \frac{q^2}{\pi} \right)  G_\Phi(\mathbf{x},\mathbf{y}) = \frac{\delta(\mathbf{x}-\mathbf{y}) }{\sqrt{g_x}} - \frac{1}{4\pi \ell^2}
~,
\end{equation}
as expected from the  action (\ref{exactSpf}).
The constant shift on the right hand side is due to the absence of the zero-mode of $\Phi$. We recognize that the solution \eqref{GF} is given by the difference between a massless and a massive scalar propagator in dS$_2$ (see also \eqref{eq:sumdiff}), along with a constant shift arising from this subtraction of the zero-modes. This particular constant shift can be thought of as a choice of parameter $\alpha$, discussed extensively in \cite{Allen:1987tz,Folacci:1992xc,Anninos:2023lin}, which arises naturally when discussing scalars in the discrete series representation of dS$_2$. As we will see (for example in \eqref{gXY}), this particular choice of $\alpha$ does not drop out in all physical observable, affecting the physics. As we vary the coupling, the relevant unitary irreducible representation of de Sitter \cite{Higuchi:1985ad,Higuchi:1986wu,Dobrev:1977qv} labeled by $\Delta$ takes values in all the possible choices.

Since $F_{12}(\mathbf{x})$ is a component of a tensor, under coordinate transformations its correlators will not transform as nicely as those of a scalar field. To simplify our analysis, we therefore define a re-scaled (scalar) field strength whose expectation values will be easier to express:
\begin{equation}
 \tilde{F}_{12}(\mathbf{x}) \equiv \frac{1}{2}\epsilon^{\mu\nu}F_{\mu\nu}= \frac{F_{12}(\mathbf{x})}{\sqrt{g_x}}\,,
\end{equation}
such that
\begin{equation}
     \langle \text{E} \lvert \tilde{F}_{12}(\mathbf{x}) \tilde{F}_{12}(\mathbf{y}) \lvert \text{E} \rangle =  \nabla^2_{\mathbf{x}}\nabla^2_{\mathbf{y}} G_\Phi(\mathbf{x},\mathbf{y}) \, .
\end{equation} 
We can compute this expectation function directly:
\begin{equation}
      \langle \text{E}| \tilde{F}_{12}(\mathbf{x}) \tilde{F}_{12}(\mathbf{y}) |\text{E}\rangle = q^2\frac{\delta(\mathbf{x}-\mathbf{y}) }{\sqrt{g_x}}-  \frac{q^4}{\pi}\frac{\Gamma(\Delta) \Gamma(1-\Delta)}{4\pi} {_2}F_1\left(\Delta,1-\Delta,1,1-\frac{u_{xy}^E}{2}\right)~.\label{chichi}
\end{equation}
 Besides the contact term, this expression is the standard scalar two-point function of a massive scalar field multiplied by $-\frac{q^4}{\pi}$, similar to the flat-space result originally derived by Schwinger \cite{Schwinger:1962tp}.

\subsection{Perturbative structure} 

The structure of expression \eqref{chichi} encodes the quantum all-loop expansion of the underlying theory. Indeed, we can systematically expand the expression at small  $\Delta(1-\Delta)=\tfrac{q^2\ell^2}{\pi}$, yielding:\footnote{To produce the expansion (\ref{gauge2ptsmallb}), and continue to arbitrary higher order, we have used the Mathematica package HypExp \cite{Huber:2005yg,Huber:2007dx}, and have ignored the contact term.}
\begin{equation}
     \langle \text{E}|\tilde{F}_{12}(\mathbf{x}) \tilde{F}_{12}(\mathbf{y}) |\text{E}\rangle {=} \frac{q^2}{4\pi\ell^2}  \left[-1+\frac{q^2\ell^2}{\pi}\left(1+\log\frac{u^E_{xy}}{2}\right)+\mathcal{O}\left(q^4 \ell^4\right)\right]~, \label{gauge2ptsmallb}
\end{equation}
and recall
\begin{equation}
u_{xy}^E\equiv 2\sin^2\frac{\Theta_{xy}}{2}~,
\end{equation}
which we defined in \eqref{eq:udef}.
The term proportional to $q^4$ stems from the one-loop vacuum polarization contribution to the gauge field propagator mediated by a virtual fermion anti-fermion pair. Continuing in this way, we arrive at a rich dataset of exact predictions for higher loop diagrams for the Schwinger model in a dS$_2$ background. Interestingly, from a K\"ahl\'en-Lehmann perspective, the one-loop correction
\begin{equation}\label{oneloopchi}
\langle \text{E}| \tilde{F}_{12}(\mathbf{x}) \tilde{F}_{12}(\mathbf{y}) |\text{E}\rangle_{\text{one-loop}} = \frac{q^4}{4\pi^2}\left(1+  \log \frac{u^E_{xy}}{2} \right)~.
\end{equation}
takes the form of a discrete series contribution described in \cite{Anninos:2023lin}.

\subsection{Late-time behavior of electric field two-point function} 

We now consider the late-time behavior, in Lorentzian signature, of the correlation functions, both when $q^2\ell^2$ is finite and when $q^2\ell^2$ is small. To do so, we must first go to Lorentzian signature by analytically continuing the $SO(3)$ invariant length \eqref{eq:udef} to the global dS$_2$ coordinates \eqref{eq:desitterdistconf} as follows
\begin{equation}
u_{xy}^E\rightarrow u^L_{xy}=\frac{\cos(T_x-T_y)-\cos(\varphi_x-\varphi_y)}{\sin T_x\,\sin T_y}~. 
\end{equation}
The equal time late-time limit, in this case, is given by $T_x=T_y\equiv T$ with $T\rightarrow 0^-$. We can take this in one of two ways. We can either take the late-time limit of \eqref{chichi} directly, after which we expand in $q^2\ell^2$, or we can work with the perturbative expansion \eqref{gauge2ptsmallb} directly. Both yield the same result. The small-$T$ expression reads (where now $\mathbf{x}$ and $\mathbf{y}$ label points on dS$_2$):
\begin{multline}\label{latechi}
     \langle \text{E}| \tilde{F}_{12}(\mathbf{x}) \tilde{F}_{12}(\mathbf{y}) |\text{E}\rangle{=}-\frac{q^4}{4\pi^2}\left\lbrace\frac{\Gamma(1-2\Delta)\Gamma(\Delta)}{\Gamma(1-\Delta)}\left[\frac{\sin^2\left(\frac{\varphi_x-\varphi_y}{2}\right)}{T^2}\right]^{-\Delta}\right.\\\left.+\frac{\Gamma(2\Delta-1)\Gamma(1-\Delta)}{\Gamma(\Delta)}\left[\frac{\sin^2\left(\frac{\varphi_x-\varphi_y}{2}\right)}{T^2}\right]^{-(1-\Delta)}+\mathcal{O}\left(T^2\right)\right\rbrace~.
\end{multline}
Taking the small $q^2\ell^2$ limit first, followed by a small $T$ limit, we find
\begin{multline}
      \langle \text{E}| \tilde{F}_{12}(\mathbf{x}) \tilde{F}_{12}(\mathbf{y}) |\text{E}\rangle {=} -\frac{q^2}{4\pi\ell^2}+\frac{q^4}{4\pi^2}\left(1+\log\frac{\sin^2\left(\frac{\varphi_x-\varphi_y}{2}\right)}{T^2}+\mathcal{O}\left(T^2\right) \right) \\  +\frac{q^6\ell^2}{4\pi^3}\left(1-\frac{\pi^2}{3}-\frac{1}{2}\left\lbrace\log\frac{\sin^2\left(\frac{\varphi_x-\varphi_y}{2}\right)}{T^2}\right\rbrace^2 + \mathcal{O}\left(T^2\right)\right)+\mathcal{O}\left(q^8\ell^4\right)~,\label{eq:latetimeloop}
\end{multline}
Here we have an explicit example where the loop-expansion results in the appearance of late-time logarithms (see for instance \cite{Ford:1984hs,Antoniadis:1985pj,Tsamis:1996qq,Polyakov:2007mm,Anninos:2014lwa,Gautier:2013aoa,LopezNacir:2016gzi}), which can be resummed to give something that ultimately decays as a de Sitter invariant power law. 

As we can deduce from \eqref{eq:latetimeloop}, each loop contribution becomes $\mathcal{O}(1)$ at exponentially-late times of order $T \sim -\exp{\left( -\frac{\pi}{2q^2\ell^2}\right)}$, signifying the breakdown of de Sitter perturbation theory once this age is reached. At least in this example, this breakdown neither signifies the breakdown of de Sitter symmetry, nor does it suggest that there are uncontrollably large effects. As we observe in the de Sitter invariant expression (\ref{latechi}), everything remains well-behaved in the infinite future. 

\section{Fermion two-point function}\label{secPsiPsi}

In this section, we consider the two-point function of the fermionic field $\Psi(\mathbf{x})$. Since we have derived an expression for the generating function of connected correlators of this theory in \eqref{exactSpf}, this particular correlation function can be found by computing 
\begin{equation}\label{eq:generatetwopoint}
 \langle \text{E}| \bar{\Psi}(\mathbf{x})  \Psi(\mathbf{y}) |E\rangle=- \left.\frac{\delta}{\delta{\eta}(\mathbf{x})}\frac{\delta}{\delta{\bar\eta}(\mathbf{y})}\log Z[\eta,\bar{\eta},J_\mu]\right\rvert_{\eta=\bar\eta=J_\mu=0}~.
\end{equation}
It follows from \eqref{eq:fermitopological} and \eqref{exactSpf}, that the $k = 0$ and $k = \pm 1$ topological sectors all contribute to this correlation function. However, it also follows from (\ref{eq:u1rot}) that this correlation function, as it stands, is not gauge invariant. To amend this, we must insert a Wilson line into the correlation function. Hence we will focus on computing the following object
\begin{equation} \label{ferGI}
\mathcal{S}_\Psi(\mathbf{x},\mathbf{y}) \equiv  \langle \text{E}| \bar{\Psi}(\mathbf{x}) \mathcal{U}(\mathbf{x},\mathbf{y}) \Psi(\mathbf{y}) |E\rangle~, \quad\quad \mathcal{U}(\mathbf{x},\mathbf{y}) \equiv e^{- i  \int_{\mathcal{C}_{{x}{y}}} ds^\mu A_\mu(s^\mu) }~,
\end{equation}
which is an appropriately dressed, and therefore gauge-invariant, object. This correlation function is implicitly labeled by the choice of path $\mathcal{C}_{{x}{y}}$,  traversed by our Wilson line, connecting the two points $\mathbf{x}$ and $\mathbf{y}$ along the two-sphere.
The generating functional of connected correlators provides us with a useful mnemonic for computing this two-point function, but we need to always remember to include this dressing factor connecting fermionic fields, to absorb any gauge dependence. 
   
Let us further denote the contribution from each topological sector by  $\mathcal{S}^{(k)}_\Psi(\mathbf{x},\mathbf{y})$ such that
\begin{equation}
\mathcal{S}_\Psi(\mathbf{x},\mathbf{y}) = \mathcal{S}^{(0)}_\Psi(\mathbf{x},\mathbf{y})+ \mathcal{S}^{(+1)}_\Psi(\mathbf{x},\mathbf{y})+\mathcal{S}^{(-1)}_\Psi(\mathbf{x},\mathbf{y})~.
\end{equation}
As we increase the number of fermionic insertions, topological sectors of larger $|k|$ will contribute. We now proceed to present the fermionic two-point function on the $S^2$ found in \cite{Jayewardena:1988td}, and then we will discuss its perturbative expansion as well as its Lorentzian dS$_2$ behavior at late times.

\subsection{A two-point function with non-vanishing trace} \label{sec:nonvanishingtrace}

The two-point function $\mathcal{S}_\Psi(\mathbf{x},\mathbf{y})$ in (\ref{ferGI})  has two (suppressed) spinorial indices. To simplify our equations, we would like to present a single function: the spinorial trace of the two-point function. Unfortunately for us, as it stands, the trace over these indices will yield a vanishing result. This is not immediately obvious, but follows from the structure of the propagator \eqref{eq:Skmaintext}. 
To find a correlation function with non-vanishing trace, we will rotate our spinors by an $SU(2)$ group element $u(\mathbf{x})$:
{
\begin{equation}\label{eq:fermrot}
\xi_{\alpha }(\mathbf{x}) \equiv \left(u^{-1}\right)_\alpha\hspace{0.1mm}^\beta{\Psi}_{\beta}(\mathbf{x}) ~, \quad\quad  \bar{\xi}^\alpha(\mathbf{x}) \equiv \bar{\Psi}^{{\beta}}(\mathbf{x}) u_{\beta}{\hspace{0.05mm}}^{\alpha}~.
\end{equation}}
Our choice of matrix $u(\mathbf{x})$ will be:
\begin{equation}\label{eq:umain}
u(\mathbf{x}) \equiv a_0(\mathbf{x})\left(\mathds{1}_{2\times 2}+i {\alpha}_i (\mathbf{x}) {\sigma}^i \right) 
\end{equation}
with 
\begin{equation}
a_0(\mathbf{x}) = \frac{\ell}{\sqrt{2(\ell^2+\mathbf{x}\cdot\mathbf{x})}}~, \qquad \alpha_1 = \frac{x^1-x^2}{\ell}~, \qquad \alpha_2 = \frac{x^1 + x^2}{\ell}~, \qquad \alpha_3 = 1~.
\end{equation}
While seemingly random, this choice of matrix $u(\mathbf{x})$ actually appears in \cref{ap:cliffordJayewardena}. It relates the Clifford matrices in the embedding space formalism to the standard tangent space construction of the $\gamma$-matrices, as can be seen in equation \eqref{ap:gammarelation} (see \cite{Jayewardena:1988td}). {Hence, correlation functions of the fermion $\xi$ would be the natural objects of interest had we used the Clifford matrices $\Gamma^\mu$ of \eqref{ap:coordinvclifford} to build our theory. As we will shortly see, this judicious choice will have enormous benefits. }

We will therefore focus on the quantity
\begin{equation}\label{eq:Sxi}
\mathcal{S}_{\xi}(\mathbf{x},\mathbf{y}) \equiv    \langle \text{E}|\bar{\xi}(\mathbf{x}) \mathcal{U}(\mathbf{x},\mathbf{y}) \xi(\mathbf{y})  | \text{E} \rangle~,
\end{equation}
and, in particular its trace: 
\begin{equation}\label{trpsipsi}
    \mathcal{T}_\xi(\mathbf{x},\mathbf{y})\equiv \text{Tr}\,\mathcal{S}_{\xi}(\mathbf{x},\mathbf{y}) ~.
\end{equation}
{More generally, we can compute the correlation functions of fermions rotated by arbitrary $SU(2)$ matrices using a simple trick. If we redefine our sources $\eta$ and $\bar\eta$ in \eqref{eq:fullzk} in the following convenient manner:
\begin{equation}
\eta(\mathbf{x})_\alpha \equiv \left( A(\mathbf{x}) \right)_\alpha\hspace{0.1mm}^\beta \eta_A(\mathbf{x})_\beta \, , \qquad \bar{\eta}(\mathbf{x})^\alpha = \bar{\eta}_B(\mathbf{x})^\beta \left( B(\mathbf{x}) \right)_\beta\hspace{0.1mm}^\alpha \, ~,
\end{equation}
then similarly to \eqref{eq:generatetwopoint} we can compute
\begin{equation}\label{eq:generalrotationofsources}
 \langle \text{E}|\left(\bar{\Psi}(\mathbf{x}) A(\mathbf{x})\right)\left(B(\mathbf{y}) \Psi(\mathbf{y})\right)|E\rangle=\left.-\frac{\delta}{\delta{\eta_A}(\mathbf{x})}\frac{\delta}{\delta{\bar{\eta}_B}(\mathbf{y})}\log Z[\eta_A,\bar{\eta}_B,J_\mu]\right\rvert_{\eta_A=\bar{\eta}_B=J_\mu=0}~.
\end{equation}
Thus to compute $\mathcal{S}_{\xi}(\mathbf{x},\mathbf{y})$ as in \eqref{eq:Sxi}, we can use \eqref{eq:generalrotationofsources} with
\begin{equation}
A(\mathbf{x}) = u(\mathbf{x}) \, , \qquad B(\mathbf{x}) = u^{-1}(\mathbf{x}) \, , 
\end{equation}
without forgetting to insert the relevant Wilson line.}

\subsection{Calculation}\label{sec:Calculation}

The two-point function (\ref{trpsipsi}) is composed of various pieces. We will derive expressions in each winding sector, starting with the zero-winding background.

\subsubsection*{Zero-winding sector} 

Following our general procedure, we take two derivatives of \eqref{exactSpf} with respect to the sources. Then the $k=0$ contribution to the fermion two-point function is found to be: 
{
\begin{equation}\label{k0int}
    \mathcal{T}^{(0)}_\xi(\mathbf{x},\mathbf{y}) = -\frac{\mathcal{N}_{\tilde{\epsilon}}^\Psi}{Z_{S^2}^{\epsilon\tilde{\epsilon}}} \int \frac{D \Phi D h}{\textnormal{vol}\mathcal{G}} J_{\Phi,h} \frac{h(\mathbf{y})}{h(\mathbf{x})} \textnormal{Tr}\left( u^{-1}(\mathbf{y}) e^{-\Phi(\mathbf{y})\gamma_*} S_{0}(\mathbf{y},\mathbf{x}) e^{-\Phi(\mathbf{x}) \gamma_*} u(\mathbf{x}) \right) \mathcal{U}(\mathbf{x},\mathbf{y})  e^{-S_\Phi} \, ,
\end{equation}}
where $\mathcal{N}^\Psi_{\tilde{\epsilon}}$ is computed in \cref{ap:integrateout} (see equation \eqref{eq:normalizationfermdet}), $Z_{S^2}^{\epsilon\tilde{\epsilon}}$ is computed in \cref{S2app} (see equation \eqref{eq:Ztotal}) and where $J_{\Phi,h}$ is the Jacobian stemming from the gauge field parametrization \eqref{Ak}, see \cref{S2app} for details. The action $S_\Phi$ can be found in \eqref{eq:Sphi} (with $k=0$) and the free fermion two-point function $S_0(\mathbf{x},\mathbf{y})$ can be read-off from \eqref{eq:Skmaintext} and is given by 
\begin{equation}
    S_0(\mathbf{x},\mathbf{y}) =  \frac{[\Omega(\mathbf{x})\Omega(\mathbf{y})]^{-\tfrac{1}{2}}}{2\pi}  \frac{\boldsymbol{\sigma}\cdot (\mathbf{x}-\mathbf{y})}{|\mathbf{x}-\mathbf{y}|^2} \, .
\end{equation}
Given our gauge choice \eqref{Ak}, the Wilson line can be expressed as:  
\begin{equation}\label{Wh}
    \mathcal{U}(\mathbf{x},\mathbf{y}) = e^{-i \int_{\mathcal{C}_{xy}} \dd z^\mu \epsilon_{\mu \nu}\partial^\nu \Phi(\mathbf{z}) + \int_{\mathcal{C}_{xy}} \dd z^\mu h(\mathbf{z})^{-1} \partial_\mu h(\mathbf{z}) } = e^{-I_{\mathcal{C}}[\Phi]} \frac{h(\mathbf{x})}{h(\mathbf{y})} \, ,
\end{equation}
where we remind the reader that the curve $\mathcal{C}_{xy}$ is chosen such that its beginning- and end-points are $\mathbf{y}$ and $\mathbf{x}$, respectively. In this expression we have also defined 
\begin{equation}\label{eq:defphiwilson}
I_{\mathcal{C}}[\Phi] \equiv   i \int_{\mathcal{C}_{xy}} \dd z^\mu \epsilon_{\mu \nu}\partial^\nu \Phi(\mathbf{z})~.
\end{equation}
The dependence on $h(\mathbf{x})$ in (\ref{k0int}) cancels against the dependence in the Wilson line (\ref{Wh}) and the corresponding path integral over the $h(\mathbf{x})$ field cancels against the factor $\text{vol}\mathcal{G}$ :
\begin{equation}
\frac{1}{\text{vol} \, \mathcal{G}}  \int  D h = \frac{1}{2\pi}~.
\end{equation}
Performing the trace over the fermionic indices, we finally arrive at
\begin{equation}\label{psipsiu}
      \mathcal{T}^{(0)}_\xi(\mathbf{x},\mathbf{y}) = +\frac{i}{2\pi \ell} \frac{\tfrac{\mathcal{N}_{\tilde{\epsilon}}^\Psi}{2\pi}}{ Z_{S^2}^{\epsilon\tilde{\epsilon}}} \int D\Phi J_{\Phi,h} e^{-\left( S_\Phi + I_{\mathcal{C}}[\Phi]\right)}\cosh\left(\Phi(\mathbf{x})-\Phi(\mathbf{y})\right) \, .
\end{equation}
Amazingly, the last remaining step is to compute a Gaussian path integral. To make this clear, let us rewrite the above equation as: 
\begin{equation}\label{eq:gaussplusj}
      \mathcal{T}^{(0)}_\xi(\mathbf{x},\mathbf{y}) = +\frac{i}{4\pi \ell} \frac{\tfrac{\mathcal{N}_{\tilde{\epsilon}}^\Psi}{2\pi}}{ Z_{S^2}^{\epsilon\tilde{\epsilon}}} \int D\Phi J_{\Phi,h}\left[e^{-\left( S_\Phi+\int \dd^2 z \sqrt{g} J_1(\mathbf{z}) \Phi(\mathbf{z})\right)} +e^{-\left( S_\Phi+\int \dd^2 z \sqrt{g} J_2(\mathbf{z}) \Phi(\mathbf{z})\right)}\right] \, ,
\end{equation}
with 
\begin{align}
J_1(\mathbf{z})&\equiv -\frac{\delta(\mathbf{z}-\mathbf{x})-\delta(\mathbf{z}-\mathbf{y})}{\sqrt{g}}-\frac{i}{\sqrt{g}}\int \dd s \frac{\dd \mathcal{C}_{xy}^\mu}{\dd s}\partial_z^\nu\left[\epsilon_{\mu\nu}\,\delta\left(z^\mu-\mathcal{C}^\mu_{xy}(s)\right)\right]~,\\
J_2(\mathbf{z})&\equiv +\frac{\delta(\mathbf{z}-\mathbf{x})-\delta(\mathbf{z}-\mathbf{y})}{\sqrt{g}}-\frac{i}{\sqrt{g}}\int \dd s \frac{\dd \mathcal{C}_{xy}^\mu}{\dd s}\partial_z^\nu\left[\epsilon_{\mu\nu}\,\delta\left(z^\mu-\mathcal{C}^\mu_{xy}(s)\right)\right]~.
\end{align}
We may now compute the Gaussian path integral in \eqref{eq:gaussplusj} by completing the square, yielding: 
\begin{equation}
      \mathcal{T}^{(0)}_\xi(\mathbf{x},\mathbf{y}) = +\frac{i}{4\pi \ell} \sum_{i=1}^2e^{ \frac{1}{2}\int \dd^2 z\, \dd^2 z' \sqrt{g_z} \sqrt{g_{z'}} \,J_i(\mathbf{z}) G_\Phi(\mathbf{z},\mathbf{z'}) J_i(\mathbf{z}')}\, ,
\end{equation}
where $G_\Phi$ is given in \eqref{GF}.
Note that the normalization factors have all dropped out of this expression.
It is quite remarkable that such an expression is even possible, given that $G_\Phi$ satisfies \eqref{eq:gphidiffeq} with the extra constant shift in the right hand side. In deriving this result, it was necessary to use the following: 
\begin{equation}
\int\dd^2x \sqrt{g}\Phi(\mathbf{x})=0~,\qquad\qquad \int\dd^2x \sqrt{g}J_1(\mathbf{x})=0~,\qquad\qquad \int\dd^2x \sqrt{g}J_2(\mathbf{x})=0~.
\end{equation}
The first of these owes to the fact that $\Phi$ is a field without a zero-mode.  

To proceed we now compute
\begin{align}
\frac{1}{2}\int \dd^2 z\, \dd^2 z' \sqrt{g_z} \sqrt{g_{z'}} &\,J_1(\mathbf{z}) G_\Phi(\mathbf{z},\mathbf{z'}) J_1(\mathbf{z}')=\nonumber\\
&+\frac{1}{2}\left[G_\Phi(\mathbf{x},\mathbf{x})+G_\Phi(\mathbf{y},\mathbf{y})-G_\Phi(\mathbf{x},\mathbf{y})-G_\Phi(\mathbf{y},\mathbf{x})\right]\nonumber\\
&-\frac{i}{2}\int_{\mathcal{C}_{xy}}\dd z^\mu \epsilon_{\mu\nu}\,\partial^\nu_z\left[G_\Phi(\mathbf{z},\mathbf{x})+G_\Phi(\mathbf{x},\mathbf{z})-G_\Phi(\mathbf{z},\mathbf{y})-G_\Phi(\mathbf{y},\mathbf{z})\right]\nonumber\\
&-\frac{1}{2}\int_{\mathcal{C}_{xy}}\dd z'^{\rho'} \epsilon_{\rho'\sigma'}\,\partial^{\sigma'}_{z'}\int_{\mathcal{C}_{xy}}\dd z^\mu \,\epsilon_{\mu\nu}\,\partial^\nu_z G_\Phi(\mathbf{z},\mathbf{z'})~.
\end{align}
The same calculation with $J_2(\mathbf{z})$ gives the complex conjugate of the above result. Therefore, we find: 
\begin{multline}
     \mathcal{T}^{(0)}_\xi(\mathbf{x},\mathbf{y}) = +\frac{i}{2\pi \ell}e^{\frac{1}{2}\left[G_\Phi(\mathbf{x},\mathbf{x})+G_\Phi(\mathbf{y},\mathbf{y})-G_\Phi(\mathbf{x},\mathbf{y})-G_\Phi(\mathbf{y},\mathbf{x})\right]-\frac{1}{2}\int_{\mathcal{C}_{xy}}\dd z'^{\rho'} \epsilon_{\rho'\sigma'}\,\partial^{\sigma'}_{z'}\int_{\mathcal{C}_{xy}}\dd z^\mu \,\epsilon_{\mu\nu}\,\partial^\nu_z G_\Phi(\mathbf{z},\mathbf{z'})}\\
     \times \cos\left(\frac{1}{2}\int_{\mathcal{C}_{xy}}\dd z^\mu \epsilon_{\mu\nu}\,\partial^\nu_z\left[G_\Phi(\mathbf{z},\mathbf{x})+G_\Phi(\mathbf{x},\mathbf{z})-G_\Phi(\mathbf{z},\mathbf{y})-G_\Phi(\mathbf{y},\mathbf{z})\right]\right)~.
\end{multline}
This is the full result for any path $\mathcal{C}_{xy}$ connecting the two endpoints $\mathbf{x}$ and $\mathbf{y}$ on $S^2$. 

We will now make a \emph{particular} choice of path $\mathcal{C}_{xy}$ in the Euclidean section, such that it follows a geodesic path (a great circle) from $\mathbf{y}$ to $\mathbf{x}$.
For example, if we fix $\mathbf{y}$ to lie on the north pole, this implies $u^E_{zy} =1+\sin(\vartheta_z)$ in the standard $S^2$ coordinates (\ref{sphere}). We can now evaluate 
\begin{equation}
\int_{\mathcal{C}_{xy}} \dd \vartheta \epsilon_{\vartheta \varphi} \partial^\varphi G_\Phi(\vartheta) = 0 \, .
\end{equation}
Thus, for this choice of dressing, we find a beautiful simplification.
Let us label the value of $G_\Phi$ at coincident points as $G_\Phi(\mathbf{x},\mathbf{x})= G_\Phi(\mathbf{y},\mathbf{y}) \equiv G_\Phi(0) $. A simple computation using \eqref{GF} yields
\begin{equation}\label{G0}
    G_\Phi(0) =\frac{1}{4} \left( \psi(1+\Delta) + \psi(2-\Delta)+2\gamma -1 \right)
\end{equation}
where $\psi(z)\equiv\frac{\Gamma'(z)}{\Gamma(z)}$ is the digamma function and $\gamma\approx 0.5772$ is the Euler-Mascheroni constant.
Putting everything together, we have (for this particular choice of dressing):
\begin{equation}
    \mathcal{T}^{(0)}_\xi(\mathbf{x},\mathbf{y}) =  \frac{i}{2\pi \ell} \exp\left( G_\Phi(0) - G_\Phi(\mathbf{x},\mathbf{y}) \right)~,
    \label{res:fermion2p_full}
\end{equation}
where we have used that $G_\Phi(\mathbf{x},\mathbf{y})= G_\Phi(\mathbf{y},\mathbf{x})$.
Later on, we will analytically continue the endpoints to Lorentzian signature. Note that the combination that appears in the exponent is independent of the procedure for removing the zero-mode of $G_\Phi$ discussed below \eqref{eq:gphidiffeq}, related to the parameter $\alpha$ in \cite{Anninos:2023lin,Allen:1987tz,Folacci:1992xc}. This will not be so in the non-trivial topological sectors.

\emph{Nota Bene:} One obvious question that arises when analytically continuing the above result to Lorentzian signature is what it implies for the Wilson line dressing computed along great circle on the $S^2$. What we believe happens is that this continues to a Wilson line that follows a complex geodesic in the Lorentzian geometry (complex geodesics for de Sitter QFT calculations were considered in \cite{Chapman:2022mqd,Aalsma:2022eru}). This is not something one would naturally compute when starting in the Lorentzian picture, but one should remember that the Euclidean state $|\text{E}\rangle$ is also not necessarily a semiclassical object from the Lorentzian perspective (see for example \cite{DiazDorronsoro:2017hti}). Below, when we analytically continue \eqref{res:fermion2p_full} to Lorentzian times, we will make this particular choice of path for our dressing in the Euclidean section.

\subsubsection*{The \texorpdfstring{$k=\pm1$}{k=plus minus 1} winding sectors} 

Proceeding as we did for the $k=0$ sectors, the $k=\pm 1$ contributions to the fermion two-point function are obtained by taking functional derivatives with respect to the sources $\eta_u(\mathbf{x})$ and $\bar\eta_{u^{-1}}(\mathbf{y})$, yielding:
\begin{multline}
\mathcal{T}^{(\pm 1)}_\xi(\mathbf{x},\mathbf{y}) = -e^{\frac{1}{2}- \frac{\pi}{2q^2 \ell^2}}\frac{\mathcal{N}_{\tilde{\epsilon}}^\Psi }{Z_{S^2}^{\epsilon\tilde{\epsilon}} } \int \frac{D \Phi D h}{\textnormal{vol}\mathcal{G}} J_{\Phi,h} \frac{h(\mathbf{y})}{h(\mathbf{x})}\\
\times\textnormal{Tr}\left(  \left(u^{-1}(\mathbf{y})\cdot e^{-\Phi(\mathbf{y})\gamma_*} \right)_\sigma\hspace{0.1mm}^\rho \chi_{\pm \rho }(\mathbf{y}) \,  \overline{\chi}_{\pm}^{\alpha}(\mathbf{x})\left(e^{-\Phi(\mathbf{x}) \gamma_*} \cdot u(\mathbf{x})\right)_\alpha\hspace{0.1mm}^\beta \, \right) \mathcal{U}(\mathbf{x},\mathbf{y}) e^{-S_\Phi} 
\end{multline}
where $\chi_{\pm}$ and $\bar{\chi}_{\pm }$ are the $k=\pm1$ zero-modes defined in \eqref{eq:zeromodes} and we have un-suppressed the spinor indices to clarify what exactly is being contracted. 

Evaluating the spinorial trace yields the following expressions: 
\begin{equation}\label{k12pt}
     \mathcal{T}^{(+1)}_\xi(\mathbf{x},\mathbf{y}) =  -\frac{\tfrac{\mathcal{N}_{\tilde{\epsilon}}^\Psi}{2\pi} }{Z_{S^2}^{\epsilon\tilde{\epsilon}} }\frac{e^{\frac{1}{2}- \frac{\pi}{2q^2 \ell^2}-W_C}{\left( \ell^2 + \bar{z}_{x} {z}_{y}\right)}}{4 \pi \ell \sqrt{\ell^2 + z_{x}\bar{z}_{x}}\sqrt{\ell^2 + z_{y}\bar{z}_{y}}}\int D \Phi J_{\Phi,h} e^{-\left( S_\Phi + I_{\mathcal{C}}[\Phi]\right)} e^{-\left(\Phi(\mathbf{x})+\Phi(\mathbf{y})\right)} 
\end{equation}
for the $k = +1$ sector, and
\begin{equation}\label{km12pt}
     \mathcal{T}^{(-1)}_\xi(\mathbf{x},\mathbf{y}) = -  \frac{\tfrac{\mathcal{N}_{\tilde{\epsilon}}^\Psi}{2\pi} }{Z_{S^2}^{\epsilon\tilde{\epsilon}} }\frac{e^{\frac{1}{2}- \frac{\pi}{2q^2 \ell^2}+ W_C}{\left( \ell^2 + {z}_{x} \bar{z}_{y}\right)}}{4 \pi \ell \sqrt{\ell^2 + z_{x}\bar{z}_{x}}\sqrt{\ell^2 + z_{y}\bar{z}_{y}}}\int D \Phi J_{\Phi,h} e^{-\left( S_\Phi + I_{\mathcal{C}}[\Phi]\right)} e^{\Phi(\mathbf{x})+\Phi(\mathbf{y})}
\end{equation}
for the $k = -1$ sector. The above expressions contain Wilson lines sensitive to the background monopole configuration, denoted as: 
\begin{equation}
 e^{W_C(\mathbf{x},\mathbf{y}) }= \exp \left[i\int_{\mathcal{C}_{xy}} \dd s^\mu \,C_\mu(\mathbf{s})\right]~,
\end{equation}
where $C_\mu$ is defined in \eqref{eq:sizeoneinstanton}.
The path integrals over $\Phi(\bf{x})$ follow as in the $k = 0$ case, by completing the square. Below we provide an answer valid when $\mathcal{C}_{xy}$ is a great circle connecting the points $\mathbf{x}$ and $\mathbf{y}$: 
\begin{equation}
    \mathcal{T}^{(+1)}_\xi(\mathbf{x},\mathbf{y}) =  - \frac{e^{\frac{1}{2}- \frac{\pi}{2q^2 \ell^2}-W_C}{\left( \ell^2 + \bar{z}_{x} {z}_{y}\right)}}{4 \pi \ell \sqrt{\ell^2 + z_{x}\bar{z}_{x}}\sqrt{\ell^2 + z_{y}\bar{z}_{y}}}e^{ G_\Phi(0) + G_\Phi(\mathbf{x},\mathbf{y})}~,
\end{equation}
\begin{equation}
    \mathcal{T}^{(-1)}_\xi(\mathbf{x},\mathbf{y}) = -  \frac{e^{\frac{1}{2}- \frac{\pi}{2q^2 \ell^2}+ W_C}{\left( \ell^2 + {z}_{x} \bar{z}_{y}\right)}}{4 \pi \ell \sqrt{\ell^2 + z_{x}\bar{z}_{x}}\sqrt{\ell^2 + z_{y}\bar{z}_{y}}}e^{ G_\Phi(0) +G_\Phi(\mathbf{x},\mathbf{y})}~.
\end{equation}
Here we have again used that $G_\Phi(\mathbf{x},\mathbf{y})= G_\Phi(\mathbf{y},\mathbf{x})$.
Note that, unlike \eqref{res:fermion2p_full} the combination that appears in the exponent above is \emph{not} independent of the $\alpha$ parameter associated to the removal of the zero-mode in $G_\Phi(\mathbf{x},\mathbf{y})$ \cite{Allen:1987tz,Folacci:1992xc,Anninos:2023lin}. In addition, notice {that the $\mathcal{T}^{(\pm 1)}_\xi$ are not expressed as pure functions of the geodesic distance $u^E_{xy}$. 
However, if we compute the absolute value $ \left\lvert \mathcal{T}^{(\pm 1)}_\xi(\mathbf{x},\mathbf{y}) \right\rvert$, we find an $SO(3)$ invariant answer:
\begin{equation}\label{gXY}
   \left\lvert \mathcal{T}^{(\pm 1)}_\xi(\mathbf{x},\mathbf{y}) \right\rvert^{2} = \frac{e^{1-\frac{\pi}{q^2 \ell^2}}}{16 \pi^2 \ell^2 } \left(1 - \frac{u^E_{xy}}{2} \right) e^{ 2[G_\Phi(0) + G_\Phi(\mathbf{x},\mathbf{y})]} \, ,
\end{equation}
which depends only on the invariant distance $u^E_{xy}$ defined in \eqref{eq:udef}. We can use this to write: 
\begin{equation}
\mathcal{T}^{(+1)}_\xi(\mathbf{x},\mathbf{y})+\mathcal{T}^{(-1)}_\xi(\mathbf{x},\mathbf{y})=-\frac{e^{\frac{1}{2}-\frac{\pi}{2q^2 \ell^2}}}{2\pi \ell } \left(1 - \frac{u^E_{xy}}{2} \right)^{\frac{1}{2}}\cos\left(\kappa(\mathbf{x},\mathbf{y})\right) e^{ G_\Phi(0) + G_\Phi(\mathbf{x},\mathbf{y})}~,\label{eq:sumoninstantonpre}
\end{equation}
where the as of yet undetermined phase $\kappa(\mathbf{x},\mathbf{y})$ is defined as follows:
\begin{equation}
e^{i\kappa(\mathbf{x},\mathbf{y})}\equiv e^{- W_C(\mathbf{x},\mathbf{y})}\frac{{\ell^2 + \bar{z}_{x} {z}_{y}}}{\left\lvert\ell^2 + \bar{z}_{x} {z}_{y} \right\rvert}~.
\end{equation}
This residual phase $\kappa(\mathbf{x},\mathbf{y})$ comes from a combination of $W_C$ evaluated along the geodesic path $\mathcal{C}_{xy}$ starting from $\mathbf{y}$ and ending at $\mathbf{x}$, as well as a multiplicative phase which depends only on these endpoints, but not on the path connecting them. Although the monopole background $C_\mu$ explicitly breaks the $SO(3)$ symmetry, the final answer yields $\kappa(\mathbf{x},\mathbf{y})=0$ if $\mathcal{C}_{xy}$ traces a great circle. The fully general computation is straightforward but tedious, so we do not reproduce it here. But we can present two simple examples.  

\paragraph{Case 1. Points along a great circle that passes through the north pole:}
In stereographic coordinates, a great circle through the north pole is represented as a straight line that goes through the origin. If we consider this configuration 
\begin{equation}
\mathbf{y}=\ell(a\,t_{\rm i}, b\, t_{\rm i})~, \qquad \mathbf{x}=\ell (a\, t_{\rm f}, b\, t_{\rm f})~, \qquad a,b \in \mathbb{R}
\end{equation}
connected by the geodesic path 
\begin{equation}
\mathbf{s}(t)=\ell (a\,t, b\, t)~, \qquad\qquad t\in (t_{\rm i},t_{\rm f})~, 
\end{equation}
then it is easy to verify that 
\begin{equation}
e^{- W_C(\mathbf{x},\mathbf{y})}=1~,\qquad \frac{{\ell^2 + \bar{z}_{x} {z}_{y}}}{\left\lvert\ell^2 + \bar{z}_{x} {z}_{y} \right\rvert}=1~, \qquad e^{i\kappa(\mathbf{x},\mathbf{y})}=1 ~.
\end{equation}

\paragraph{Case 2. Points along the equator:} We can perform a second straightforward check for $\mathbf{x}$ and $\mathbf{y}$ along the equator. In stereographic coordinates, points satisfying $\mathbf{x}\cdot\mathbf{x}=\mathbf{y}\cdot\mathbf{y}=\ell^2$ are situated along the equator, as can be deduced from \eqref{coordstereo}. Now consider the configuration 
\begin{equation}
\mathbf{y}=\ell(\cos t_{\rm i}, \sin t_{\rm i})~, \qquad \mathbf{x}=\ell(\cos t_{\rm f}, \sin t_{\rm f})~,
\end{equation}
connected by the geodesic path 
\begin{equation}
\mathbf{s}(t)=\ell (\cos t, \sin t)~, \qquad\qquad t\in (t_{\rm i},t_{\rm f})~, 
\end{equation}
then a straightforward computation yields 
\begin{equation}
e^{- W_C(\mathbf{x},\mathbf{y})}=e^{\frac{i}{2}(t_{\rm f}-t_{\rm i})}~,\qquad \frac{{\ell^2 + \bar{z}_{x} {z}_{y}}}{\left\lvert\ell^2 + \bar{z}_{x} {z}_{y} \right\rvert}=e^{-\frac{i}{2}(t_{\rm f}-t_{\rm i})}~, \qquad e^{i\kappa(\mathbf{x},\mathbf{y})}=1 ~.
\end{equation}
To show that the phase vanishes for any geodesic path connecting arbitrary endpoints, it suffices to consider a general $SO(3)$ rotation of the above equatorial paths. Although the general case yields more cumbersome intermediate expressions, we have checked that the result is always $\kappa(\mathbf{x},\mathbf{y})=0$.

\bigskip
Thus, having dealt with the phase, we can finally write: 
\begin{equation}
\mathcal{T}^{(+1)}_\xi(\mathbf{x},\mathbf{y})+\mathcal{T}^{(-1)}_\xi(\mathbf{x},\mathbf{y})=-\frac{e^{\frac{1}{2}-\frac{\pi}{2q^2 \ell^2}}}{2\pi \ell } \left(1 - \frac{u^E_{xy}}{2} \right)^{\frac{1}{2}} e^{ G_\Phi(0) + G_\Phi(\mathbf{x},\mathbf{y})}~,\label{eq:sumoninstanton}
\end{equation}
 By now we have derived a fully non-perturbative result for the trace of the two-point function of the fermion $\xi$. Noting the prefactor in the above expressions, at weak coupling the contributions $ \mathcal{T}^{(\pm 1)}_\xi(\mathbf{x},\mathbf{y})$ from the higher winding sectors are exponentially suppressed, owing to their non-perturbative nature. One way to interpret this effect is to think of a highly suppressed fluctuation creating an instanton that mediates an interaction between the pair of fermions. This physics \emph{has no counterpart in the Schwinger model in flat space}, since the $\ell\rightarrow\infty$ limit removes the suppression. An analogous effect can be found in the finite temperature Schwinger model  on a spatial $S^1$ \cite{Sachs:1991en}, and the obvious connection between these two results exemplifies the thermal nature of the Euclidean de Sitter vacuum \cite{Figari:1975km,Gibbons:1977mu}.

\subsubsection*{Final answer} 
{We can now combine everything and write down the trace of the fermion two-point function connected by a geodesic Wilson line, in all its non-perturbative glory:
\begin{equation}
\mathcal{T}_\xi(\mathbf{x},\mathbf{y})=i\frac{e^{ G_\Phi(0)- G_\Phi(\mathbf{x},\mathbf{y})}}{2\pi\ell}\left[ 1 + i\,e^{\frac{1}{2}-\frac{\pi}{2q^2 \ell^2}}\left(1 - \frac{u^E_{xy}}{2} \right)^{\frac{1}{2}} e^{2G_\Phi(\mathbf{x},\mathbf{y})}\right]~.
\end{equation}
This simple function captures a lot of interesting physics about de Sitter, as we will analyze below. 
Incredibly, although one might have imagined the Wilson line (or the explicit monopole backgrounds) spoiling the $SO(3)$ symmetry, we find instead that the final result $\mathcal{T}_\xi(\mathbf{x},\mathbf{y})$ only depends on the invariant distance $u_{xy}^E$ so long as the Wilson line $W_C$ traces a great circle on the $S^2$. We have therefore ascertained that the $SO(3)$ symmetry survives unscathed. As previously mentioned, the Lorentzian continuation of this Wilson line configuration is somewhat devoid of a semiclassical meaning, but we nevertheless take solace in the fact that the $SO(3)$ symmetry (and hence the $SO(1,2)$ symmetry in Lorentzian signature) continues to hold non-perturbatively.}

\subsection{Perturbative structure of the fermion two-point function} 

In the perturbative regime  $q^2\ell^2\sim 0$, we are only sensitive to the $k=0$ sector. We can expand \eqref{res:fermion2p_full} perturbatively at small $q^2\ell^2$:
\begin{equation}\label{loopsPsi}
    \mathcal{T}_\xi(\mathbf{x},\mathbf{y})= \frac{i }{2\pi \ell } \left[  1 + \frac{q^2\ell^2}{24 \pi} \left( \pi^2-6 \textnormal{Li}_2\left(1-\frac{u^E_{xy}}{2}\right) \right) + \mathcal{O}\left(q^4 \ell^4\right) \right]~.
\end{equation}
In the flat space Schwinger model, the papers \cite{Adam:1993fy,Adam:1996qm} explored the relation between the small $q^2$ expansion of the fully non-perturbative fermion two-point function and compared it to the Feynman diagrammatic expansion. We would like to similarly retrieve (\ref{loopsPsi}) via a Feynman diagram calculation on the $S^2$. To do this we need to introduce our perturbative propagators.

\textbf{Photon propagator:} Working in Lorenz gauge (\ref{Ak}), a small gauge field fluctuations is expressed as $A_\mu(\mathbf{x}) = \epsilon_{\mu\nu} \partial^\nu \Phi(\mathbf{x})$. Notably, in this gauge the Fadeev-Popov ghosts decouple. The perturbative propagator of $\Phi(\mathbf{x})$ in position space $G^{\rm pert}_\Phi(\mathbf{x},\mathbf{y})$ satisfies (the $q\rightarrow0$ limit of \eqref{eq:gphidiffeq}):
\begin{equation}
 \frac{1}{q^2} \nabla^4 G^{\rm pert}_\Phi(\mathbf{x},\mathbf{y}) = \frac{\delta(\mathbf{x}-\mathbf{y}) }{\sqrt{g_x}} - \frac{1}{4\pi \ell^2}
~,
\end{equation}
and reads,
\begin{equation}\label{Gphi}
G^{\rm pert}_\Phi(\mathbf{x},\mathbf{y}) =   \frac{q^2\ell^2}{{24 \pi}} \left( 6 \textnormal{Li}_2\left(1-\frac{u^E_{xy}}{2}\right) + 6 - \pi^2 \right)~.
\end{equation}
which is the expression \eqref{GF} expanded to lowest order in $q^2\ell^2$.

The coincident point limit of (\ref{Gphi}), whereby $u^E_{xy} \to 0$, agrees with the flat space counterpart reported, for instance, in \cite{Adam:1993fy,Adam:1996qm}. From $G^{\rm pert}_\Phi(\mathbf{x},\mathbf{y})$ we can readily compute the photon propagator in Lorenz gauge, which reads
\begin{equation}\label{photonprop}
D_{\mu\nu}(\mathbf{x},\mathbf{y}) = \epsilon_{\mu\rho}(\mathbf{x})  \epsilon_{\nu\lambda}(\mathbf{y})\partial^\rho_x\,\partial^\lambda_y\, G_\Phi^{\rm pert}(\mathbf{x},\mathbf{y})~.
\end{equation}
We can read off the interaction vertex $i \gamma^\mu  A_\mu \bar{\Psi} \Psi$ from the action \eqref{eq:Sschw}, which governs the structure of all diagrams. Since this is a cubic vertex, the one-loop contribution comes from the diagram as in figure \ref{fig:loopdiagram}.

\textbf{Fermion propagator:} The fermion propagator requires much less introduction, it satisfies
\begin{equation}
    \gamma^\mu{\nabla}_\mu S_0(\mathbf{x},\mathbf{y}) =\frac{\delta(\mathbf{x}-\mathbf{y})}{\sqrt{g}} \mathds{1}_{2\times2}~.
\end{equation}
and has already appeared previously in equation \eqref{eq:Skmaintext} with $k=0$. In sterographic coordinates it reads 
\begin{equation}
S_0(\mathbf{x}, \mathbf{y})=\frac{\left[\Omega(\mathbf{x})\Omega(\mathbf{y})\right]^{-\tfrac{1}{2}}}{2\pi}  \frac{\boldsymbol{\sigma}\cdot (\mathbf{x}-\mathbf{y})}{|\mathbf{x}-\mathbf{y}|^2}=\left[\Omega(\mathbf{x})\Omega(\mathbf{y})\right]^{-\tfrac{1}{2}}S^{\rm flat}(\mathbf{x},\mathbf{y})
\end{equation}
where $S^{\rm flat}(\mathbf{x},\mathbf{y})$ is the flat space fermion propagator whose explicit form is given in \eqref{sFlat}.

\textbf{Loop calculation:} The order $q^2\ell^2$ term is a one-loop contribution to the fermion propagator. In our conventions, the gauge coupling $q$ appears only in the gauge-kinetic term. Thus there no factors of $q$ are associated with interaction vertices. Rather, photon propagators will carry all the explicit factors of $q^2\ell^2$, as can be deduced from \eqref{Gphi}. We can now write down our position space loop integral computing the diagram in figure \ref{fig:loopdiagram}:
\begin{multline}\label{eq:fermiononeloop}
    \text{Tr} \,   \langle  \text{E}| \bar{\xi}({\bf x}) {\xi}({\bf y})|\text{E} \rangle_{\text{one-loop}} =2\times\frac{i^2}{2!}\int \dd^2 \mathbf{x}' \dd^2 \mathbf{y}' \sqrt{g_{x'}} \sqrt{g_{y'}} \\  \textnormal{Tr} \left(u^{-1}(\mathbf{y})S_0(\mathbf{y},\mathbf{x}') \gamma^\mu(\mathbf{x}') D_{\mu \nu} (\mathbf{x}',\mathbf{y}') S_0(\mathbf{y}',\mathbf{x}') \gamma^\nu(\mathbf{y}') S_0(\mathbf{y}', \mathbf{x})u(\mathbf{x})\right) \, .
\end{multline}
The factor of $2!$ comes from expanding the path integral to this order, and the extra factor of 2 comes from considering the number of possible contractions to form this diagram. Note that at this order in perturbation theory, we have not included any Wilson line dressing between the external fermionic operators. 

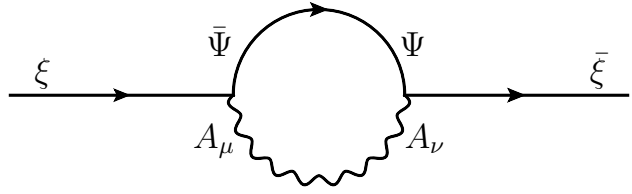
\begin{figure}[t!]
    \centering
    \includegraphics[width=0.5\textwidth]{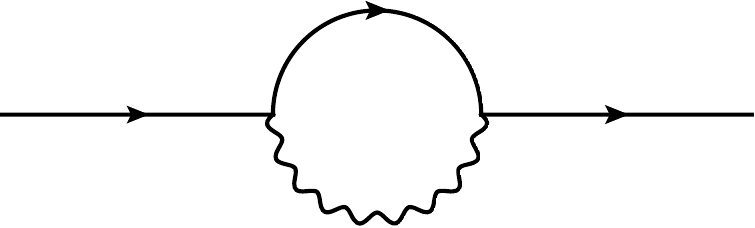}
    \put(-225,40){$\mathlarger{\mathlarger{{\xi}}}$}
    \put(-15,40){$\mathlarger{\mathlarger{\bar{\xi}}}$}
     \put(-165,15){$\mathlarger{\mathlarger{A_\mu}}$}
     \put(-85,15){$\mathlarger{\mathlarger{A_\nu}}$}
      \put(-87,50){$\mathlarger{\mathlarger{{\Psi}}}$}
     \put(-160,50){$\mathlarger{\mathlarger{\bar{\Psi}}}$}
    \caption{One-loop diagram contributing to the fermion propagator. Recall that $
    \xi$ and $\bar{\xi}$ are related to the bare fermion via an $SU(2)$ rotation as in \eqref{eq:fermrot}.}
    \label{fig:loopdiagram}
\end{figure}
Here we find a case where working in the stereographic coordinates of \eqref{spherestereo} results in a simplification of the integrand. 
Using \eqref{ap:2dcommute}, as well as the explicit form of the $\gamma$-matrices in these coordinates, we can massage this expression to:
\begin{multline}
     \text{Tr} \,   \langle  \text{E}| \bar{\xi}({\bf x}) {\xi}({\bf y})|\text{E} \rangle_{\text{one-loop}} =2\times \frac{i^2}{2!}[\Omega(\mathbf{x})\Omega(\mathbf{y})]^{-1/2}  \int \dd^2 \mathbf{x'} \dd^2 \mathbf{y'} \\  
    \textnormal{Tr} \left(u^{-1}(\mathbf{y}) S^{\rm flat}(\mathbf{y},\mathbf{x'}) \gamma_* \gamma^{a}\left[{\partial}^{x'}_{a} \partial_{b}^{y'} G^{\rm pert}_\Phi(\mathbf{x'},\mathbf{y'})\right] S^{\rm flat}(\mathbf{y}',\mathbf{x}') \gamma^{b} \gamma_* S^{\rm flat}(\mathbf{y'}, \mathbf{x})  u(\mathbf{x}) \right) \, ,
    \label{fermionFeyn}
\end{multline}
where we remind the reader that $\gamma$-matrices with latin indices are evaluated in the tangent frame \eqref{eq:localframegamma} and therefore have no coordinate dependence. Surprisingly, in these coordinates, the integral is only very mildly sensitive to the fact that we are on $S^2$. The only reference to the $S^2$ geometry comes form the functional form of $G_\Phi^{\rm pert}$. One can check, by inspection of the integrand, that the above expression is ultraviolet and infrared finite. In another remarkable feat, the equivalent flat space expression would vanish identically \cite{Adam:1993fy,Adam:1996qm}, but this is not the case on the $S^2$. 

We can compute this integral analytically. Doing so simply requires integration by parts and making use of \eqref{eq:Sflatdiffeq}. The result is: 
\begin{equation}
\text{Tr} \,   \langle  \text{E}| \bar{\xi}({\bf x}) {\xi}({\bf y})|\text{E} \rangle_{\text{one-loop}}=\frac{i}{2\pi\ell}\left[G^{\rm pert}_\Phi(0)-G^{\rm pert}_\Phi(\mathbf{x},\mathbf{y})\right]~.
\end{equation}
One can check that this exactly matches the one-loop prediction in (\ref{loopsPsi}), which reads
\begin{equation}
\mathcal{T}_\xi(\mathbf{x},\mathbf{y})^{\rm one-loop}= i   \frac{q^2\ell}{48 \pi^2} \left(\pi^2- 6 \textnormal{Li}_2\left(1-\frac{u^E_{xy}}{2}\right) \right) ~.
\label{GF1loop}
\end{equation}

As a sanity check, we also compute this integral numerically, on a particular slice in position space. We use rotational invariance to move the external point $\mathbf{x}$ to the origin and further set one of the two coordinates in $\mathbf{y}$, namely $y^1$, to zero. The results are displayed in figure \ref{fig:plot}, which demonstrate beautiful agreement. Additionally, there is little obstructing us from proceeding in this way at higher-loop orders. 

Loop expressions such as \eqref{eq:fermiononeloop} often appear in perturbative de Sitter QFT computations. These expressions are generally very cumbersome to compute. There is a slew of recent literature developing methods for computing loop diagrams in de Sitter, including \cite{Marolf:2010zp,Anninos:2014lwa,Muhlmann:2022duj,DiPietro:2023inn,Arkani-Hamed:2023kig,Chakraborty:2023qbp,Qin:2023bjk,Cacciatori:2024zrv,Heckelbacher:2020nue,Heckelbacher:2022hbq}. In our case, we benefit in that we can explicitly test these methods against an exact one-loop expression (\ref{loopsPsi}) obtained from a non-perturbative result.
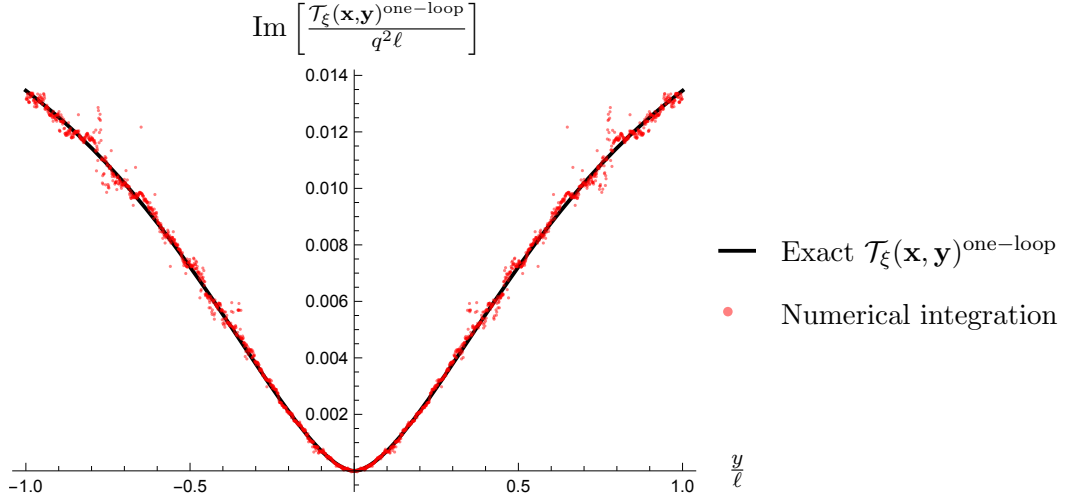
\begin{figure}[t!]
    \centering
    \includegraphics{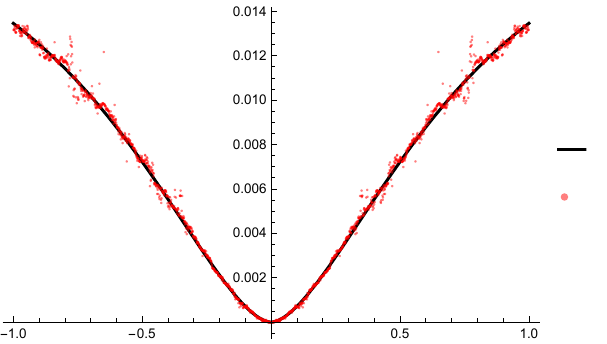}
    \put(0,90){Exact $\mathcal{T}_\xi(\mathbf{x},\mathbf{y})^{\rm one-loop}$ }
    \put(0,66){Numerical integration}
    \put(-200,175){$\text{Im}\left[\frac{\mathcal{T}_\xi(\mathbf{x},\mathbf{y})^{\rm one-loop}}{ q^2\ell}\right]$}
    \put(-20,8){$\frac{y}{\ell}$}
    \caption{One-loop contribution to the fermion propagator (\ref{res:fermion2p_full}) evaluated at $\mathbf{x}=(0,0)$ and $\mathbf{y}=(0,y)$. The thick line corresponds to the exact one-loop contribution (\ref{GF1loop}) while the red points correspond to the numerical evaluation of (\ref{fermionFeyn}).}
    \label{fig:plot}
\end{figure}

\subsection{Late-time behavior of the exact fermion two-point function}
Given that we have derived an exact, all-orders, two-point function for the charged fermions in this theory, we can now analyze the behavior of this correlation function in the Lorentzian continuation to de Sitter, at late times. Just as we did for the gauge field two-point function, we obtain the Lorentzian behavior by replacing $u_{xy}^E\rightarrow u_{xy}^L$ with $u_{xy}^L$ found in \eqref{eq:desitterdistconf}. Moreover we will restrict to equal-time insertions $T_x=T_y\equiv T$ and take the late time limit $T\rightarrow 0^-$. 

We first expand the answer in the zero-instanton sector, given by  (\ref{res:fermion2p_full}), and find:
\begin{multline}
    \mathcal{T}^{(0)}_\xi(\mathbf{x},\mathbf{y}) {=} \frac{i}{2\pi \ell}e^{\frac{1}{4}\left(2\gamma+\psi(\Delta)+\psi(1-\Delta)\right)} \left( \frac{\sin^2\left( \frac{\varphi_x - \varphi_y}{2} \right)}{T^2} \right)^{\frac{1}{4}} \\ \times  \exp\left[ \frac{\Gamma(\Delta) \Gamma(1-2\Delta)}{4 \Gamma(1-\Delta)} \left( \frac{\sin^2\left( \frac{\varphi_x - \varphi_y}{2}\right)}{T^2} \right)^{-\Delta} + (\Delta \leftrightarrow 1-\Delta )+ \mathcal{O}\left(T^2\right) \right]~,\label{latetimek0}
\end{multline}
where now $\mathbf{x}$ and $\mathbf{y}$ label points on dS$_2$ and $\varphi_x$ and $\varphi_y$ are points on the spatial $S^1$ at fixed $T$ and $\psi(x)$ is again the digamma function. We note that the expression grows with time, indicating a growing correlation as the two fermions are more separated on the $S^1$. This is an intrinsically de Sitter effect, with no flat space counterpart given the vanishing of the trace in flat space. This is perhaps a result of the confining nature of this theory, with the fermions connected by an Abelian flux tube. 

The contribution from the 1-instanton sector is given in \eqref{eq:sumoninstanton}. If we compute the late-time behavior, we find
\begin{multline}
    \left(\mathcal{T}^{(+ 1)}_\xi(\mathbf{x},\mathbf{y})  +\mathcal{T}^{(- 1)}_\xi(\mathbf{x},\mathbf{y}) \right) {=} -  \frac{i}{2\pi \ell} \,  e^{\frac{1}{4}\left(2\gamma+\psi(\Delta)+\psi(1-\Delta)\right)} \,  \left( \frac{\sin^2\left( \frac{\varphi_x - \varphi_y}{2} \right)}{T^2} \right)^{\frac{1}{4}} \\ \times  \exp\left[ -\left\lbrace \frac{\Gamma(\Delta) \Gamma(1-2\Delta)}{4 \Gamma(1-\Delta)} \left( \frac{\sin^2\left( \frac{\varphi_x - \varphi_y}{2}\right)}{T^2} \right)^{-\Delta} + (\Delta \leftrightarrow 1-\Delta )\right\rbrace+ \mathcal{O}\left(T^2\right) \right]~,\label{latetimetop}
\end{multline}
which is a truly remarkable result. Although at early times and at small $q\ell$, these non-zero instanton sectors are non-perturbatively suppressed relative to the $k=0$ sector, the late-time limit conspires such that all topological sectors contribute equivalently. 
Compare for a moment \eqref{latetimek0} with \eqref{latetimetop}, and you will conclude that it is impossible, without prior knowledge, to tell which term is leading in the small $q^2\ell^2$ limit. Late-time physics in de Sitter is bound to confound, all the more in this exactly-solvable example, where the late-time limit makes a mockery of perturbation theory. Even without gravity, we have found a situation where the late-time limit forces us to consider a situation where non-perturbative saddles necessarily compete with the leading semi-classical contribution---much like the replica wormhole story in the Euclidean analysis of old black holes \cite{Almheiri:2019qdq}.

If such things can happen generically in de Sitter, we must carefully assess what a perturbative approach to cosmological correlators will capture of the complete nature of cosmological observables. It is thus incumbent upon us to both analyze this model as much as possible, and discover new exactly solvable models on a rigid de Sitter background.

\section{Outlook}\label{outlook}

That the Schwinger model on a rigid dS$_2$ geometry has retained its exact solvability means we now have been provided with a valuable model to study quantum features in de Sitter space. It is not clear what the general lesson is. Can any exactly-solvable theory in flat space be placed on an arbitrary maximally symmetric spacetime and still retain its solvability? It is crucial that we take this question seriously, at least for two spacetime dimensions  (for example, the Schwinger model is also solvable on the Poincar\'e disk \cite{Sachs:1991en}). The Schwinger model on dS$_2$ exhibits rich phenomena of important physical content, which can serve to test the various tools and methods developed over the last few years to study de Sitter quantum field theory in and beyond the perturbative limit. Solvable models of de Sitter, both field-theoretic and gravitational \cite{Martinec:2014uva,Anninos:2021ene,Muhlmann:2022duj,Anninos:2017eib,Coleman:2021nor,Batra:2024kjl,Anninos:2017hhn,Maldacena:2019cbz,Cotler:2024xzz,Anous:2020nxu,Pethybridge:2024qci,Anninos:2020cwo,Narovlansky:2023lfz,Susskind:2022bia,Castro:2023dxp,Castro:2023bvo} will be crucial tools in our pursuit to crack this problem open. 

The fundamental matter fields of this theory are fermions. Many papers on de Sitter have instead focused on scalar field correlation functions, for their simplicity. Recent exceptions include \cite{Letsios:2020twa,Letsios:2022slc,Letsios:2023awz,Letsios:2024nmf,Pethybridge:2021rwf,Schaub:2023scu,Anninos:2023exn}. Our theory at hand admits such a scalar operator, but it is composite. If we define:
\begin{equation}
\phi(\mathbf{x}) = \lim_{\mathbf{y}\to\mathbf{x}} \text{Tr}\,\bar{\Psi}(\mathbf{x})\mathcal{U}(\mathbf{x},\mathbf{y})\Psi(\mathbf{y})~,
\end{equation}
by a normal-ordering procedure, then we note that we have a well-defined bosonic operator which can be bootstrapped \cite{Strominger:2001pn,Maldacena:2002vr,Bzowski:2013sza,Anninos:2014lwa,Anninos:2014hia,Arkani-Hamed:2017fdk,Arkani-Hamed:2018kmz,Jazayeri:2021fvk,Sleight:2020obc,Hogervorst:2021uvp,DiPietro:2021sjt}. Moreover, since it is evaluated at coincident points, the dependence on the Wilson line drops out.
The two-point function of $\phi(\mathbf{x})$ can be obtained by taking limits of the $\Psi(\mathbf{x})$ four-point function. Since we have the generating functional of connected correlators, we can compute this exactly (although we will refrain from doing so here) and the answer receives contributions from the $k = 0,\pm1$ and $k=\pm2$ instanton sectors. In \cite{Jayewardena:1988td}, the two-point function of a slightly different scalar $\phi_\xi(\mathbf{x})\equiv\lim_{\mathbf{y}\to\mathbf{x}} \text{Tr}\,\bar{\xi}(\mathbf{x})\mathcal{U}(\mathbf{x},\mathbf{y})\xi(\mathbf{y})$ was analyzed, and we will quote certain parts of the result here, for discussion. The $k=0$ contribution is given by
\begin{equation}
 \langle \text{E} | \phi_\xi(\mathbf{x}) \phi_\xi(\mathbf{y}) | E\rangle^{(0)} = -\frac{1}{4\pi^2\ell^2}\left(1-\frac{1}{u^E_{xy}}e^{4 (G_\Phi(0)-G_\Phi(\mathbf{x},\mathbf{y}))} \right)~.
\end{equation}
Going to Lorentzian signature and placing the operators on a late-time slice, the above correlator tends to a constant as the slice is pushed to $\mathcal{I}^+$. The $k=\pm1$ contributions sum to a constant, so we will omit writing them here, but as before, they are exponentially suppressed by a factor $e^{-\frac{\pi}{2q^2\ell^2}}$. The $k=\pm2$ contribution are simpler to write down, and \cite{Jayewardena:1988td} finds:
\begin{equation}
\sum_{k=\pm 2}  \langle \text{E} | \phi_\xi(\mathbf{x})\phi_\xi(\mathbf{y}) | E\rangle^{(k)} =  \frac{u^E_{xy}}{16\pi^2\ell^2} e^{2-\frac{2\pi}{q^2\ell^2} + 4  (G(0)+G(\mathbf{x},\mathbf{y}))}~.
\end{equation}
Going to Lorentzian signature, and taking the equal late-time limit, the above $k=\pm 2$ contributions decay to zero. As such, the late-time structure of the whole two-point function of $ \langle \text{E}|\phi_\xi(\mathbf{x})\phi_\xi(\mathbf{y})|\text{E}\rangle$ tends to a constant in the late-time limit. This is consistent with cluster decomposition, since the scalar $\phi_\xi(\mathbf{x})$ acquires a vacuum expectation value, owing to the formation of a chiral condensate. In future work we will analyze the four-point function of $\phi(\mathbf{x})$, since it is exactly computable thanks to the Gaussian nature of this theory. We anticipate it will yield  interesting lessons for the de Sitter bootstrap in two-dimensions.

We end this paper by noting the non-local nature of the fermionic observables of this theory. We view this as a toy version of the situation one encounters when gravity is dynamical in de Sitter, due to the constraints that must be imposed by diffeomorphism invariance. Already in this toy example, we see that this non-locality can lead to growing correlations, as in \eqref{latetimek0}, that, at least na\"ively, seem to defy the standard logic of cluster decomposition \cite{Anninos:2011kh}. Related to this, it might be of particular interest to couple this model to a weakly fluctuating metric to learn what the fate of these growing correlations is.

\section*{Acknowledgements}

It is a pleasure to acknowledge Atsushi Higuchi for reinvigorating our interest in solvable QFTs in de Sitter and Christoph Adam for generously sharing his PhD thesis with us. We also thank Paolo Benincasa, Pietro Benetti Genolini, Vasileios Letsios, Manuel Loparco, Beatrix M\"uhlmann, Guilherme Pimentel, Ben Pethybridge,  Jiaxin Qiao, Vladimir Schaub, Zimo Sun, and Kamran Salehi Vaziri for useful discussions. We would also like to thank Johnny Brendan Gleeson for pointing out important sign mistakes in an earlier draft of this paper. D.A. is funded by the Royal Society under the grant ``Concrete Calculables in Quantum de Sitter” and the STFC Consolidated grant ST/X000753/1. T.A. is supported by the UKRI Future Leaders Fellowship ``The materials approach to quantum spacetime'' under reference MR/X034453/1.  A.R.F. is funded by the Royal Society under the grant ``Boundaries and Defects in Quantum Field Theory and Gravity.''

\appendix 

\section{Conventions} \label{ap:conventions}

\subsection{Spinor Conventions}\label{ap:SpinorConventions}
Throughout this paper, we work in 2-dimensional \emph{Euclidean} spacetime and follow the conventions of \cite{Freedman:2012zz} (see also Chapter 12 of \cite{Green:2012pqa}). In order to define Dirac spinors, we must first introduce the tangent bundle on the spacetime. For a general $n$-dimensional Euclidean manifold with metric $g_{\mu\nu}$, there is a local tangent frame $\delta_{ab}$ which represents a copy of $\mathbb{R}^n$ at each point. We will generally denote this local frame using latin $a,b,c,\dots$, or hatted indices $\hat{\mu}, \hat{\nu},\dots$. An important auxiliary quantity is the frame field or zweibein $e_\mu^a$, which serves as a map from the tangent space into the manifold: 
\begin{equation}
g_{\mu\nu}=e^a_\mu e^b_\nu\, \delta_{ab}~,\qquad\qquad g^{\mu\nu}=e_a^\mu e_b^\nu\, \delta^{ab}~.
\end{equation}
Greek indices are raised and lowered by $g^{\mu\nu}$ and $g_{\mu\nu}$ respectively, and latin indices are raised and lowered by $\delta^{ab}$ and $\delta_{ab}$ respectively.

To define a spinor-covariant derivative on the two-dimensional manifold, we first need a \emph{Clifford algebra}, i.e. a set of matrices satisfying: 
\begin{equation}\label{eq:clifford}
\{\gamma^\mu,\gamma^\nu\}=2g^{\mu \nu}\,\mathds{1}_{2\times 2}~.
\end{equation}
If we define: 
\begin{equation}\label{eq:localframegamma}
\gamma^{\hat{1}}=\sigma^1~,\quad\quad \gamma^{\hat{2}}=\sigma^2~, \qquad\qquad \gamma^\mu\equiv e^\mu_a \gamma^a~,
\end{equation}
we can easily verify that \eqref{eq:clifford} is satisfied. {Although we will often suppress spinor indices in this paper, let us take a brief moment to specify our spinor-index conventions. The fundamental fermion $\Psi_\alpha$ will transform in the $\boldsymbol{2}$ of $SO(3)\cong SU(2)$ while the conjugate spinor $\bar{\Psi}^\beta$ will transform in the $\bar{\boldsymbol{2}}$ representation. To distinguish between these two representations we will, respectively, denote the spinor's index as down or up.  Clifford algebra elements thus have a mixed index structure, with one index down and one index up:
\begin{equation}
\left(\gamma^{\mu}\right)_\alpha\hspace{0.1mm}^\beta
\end{equation}
and act on spinors $\Psi$ by left multiplication, whereas they act on conjugate spinors $\bar\Psi$ by right multiplication.
}

There are two more elements of the Clifford algebra in two-dimensions that are needed. First is the generator of rotations and translations of the tangent space, in the spin-$\tfrac{1}{2}$ representation: 
\begin{equation}
\gamma^{ab}\equiv\frac{1}{2}\left[\gamma^a, \gamma^b\right]~.
\end{equation}
The second is the highest element of the Clifford algebra:\footnote{This is sometimes denoted as $\gamma_5$ in four-dimensions.}
\begin{equation}\label{gammast}
\gamma_*\equiv -i\gamma_{\hat{1}}\gamma_{\hat{2}}= \sigma^3~.
\end{equation}

Now we are ready to define covariant derivatives of spinors: 
\begin{equation}
 \nabla_\mu\Psi\equiv \left(\partial_\mu +\frac{1}{4}\omega_{\mu a b}\gamma^{ab}\right)\Psi~,
\end{equation}
where the \emph{spin-connection} satisfies: 
\begin{equation}
\omega_\mu^{~~~~~~~a b}=\frac{1}{2}e^{\nu a}\left(\partial_\mu e_\nu^b-\partial_\nu e_\mu^b\right)-\frac{1}{2}e^{\nu b}\left(\partial_\mu e_\nu^a-\partial_\nu e_\mu^a\right)-\frac{1}{2}e^{\rho a}e^{\sigma b}\left(\partial_\rho e_{\sigma c}-\partial_\sigma e_{\rho c}\right)e^c_\mu~.\label{eq:spincon}
\end{equation}
This definition ensures that under \emph{local} rotations of the tangent space $\Psi(x)\rightarrow e^{-\frac{1}{4}\lambda^{ab}(x)\gamma_{ab}}\Psi(x)$, then the covariant derivative of $\Psi(x)$ transforms as a spinor as well $\nabla_\mu\Psi(x)\rightarrow e^{-\frac{1}{4}\lambda^{ab}(x)\gamma_{ab}}\nabla_\mu\Psi(x)$~.

The introduction of the element $\gamma_*$ allows us to define the projectors: 
\begin{equation}
    P_L=\frac{1+\gamma_*}{2}~,\qquad\qquad\qquad P_R=\frac{1-\gamma_*}{2}~,
\end{equation}
where the subscripts $L, R$ harken to their interpretation as projectors onto left- and right-chirality spinors in Lorentzian signature. 

To go to Lorentzian signature, starting from the Euclidean picture above, we make the following changes: 
\begin{equation}
x^2\rightarrow i x^0           
\end{equation}
along with the following slight modification of the Clifford algebra: 
\begin{equation}
\gamma^{\hat{0}}=i\sigma^2~,\quad\quad \gamma^{\hat{1}}=\sigma^1~,\qquad\qquad \gamma_*\equiv -\gamma_{\hat{0}}\gamma_{\hat{1}}= \sigma^3~.
\end{equation}
In Euclidean signature, the spinors $\bar\Psi$ and $\Psi$ are independent, however, in Lorentzian they are related as follows: 
\begin{equation}
\bar\Psi\equiv i\Psi^\dagger \gamma^{\hat{0}}~.
\end{equation}

\subsection{Geometry conventions}\label{ap:Geometry}

The Euclidean geometry prominently used in this paper is the $S^2$ which can be described by its embedding in $\mathbb{R}^3$. Throughout this paper, we will denote vectors in $\mathbb{R}^3$ with an arrow, such as $\vec{\mathbf{r}}\in \mathbb{R}^3$, while vectors in $\mathbb{R}^2$ will be denoted without an arrow, e.g. $\mathbf{x}\in \mathbb{R}^2$~. The $S^2$ is often represented by its embedding formula: 
\begin{equation}
\vec{\mathbf{r}}\cdot{\vec{\mathbf{r}}}=\ell^2~.
\end{equation} 
Distances are computed using the $SO(3)$ invariant distance: 
\begin{equation}
\cos\Theta_{xy}\equiv \frac{\vec{\mathbf{r}}_x\cdot\vec{\mathbf{r}}_y}{\ell^2}~, \quad\quad\quad u_{xy}^E\equiv 1-\cos\Theta_{xy}~.
\end{equation}

Most commonly, we will work in stereographic coordinates:
\begin{equation}
\vec{\mathbf{r}} = \frac{2\ell^2}{\ell^2 +\mathbf{x}\cdot \mathbf{x}} \left( x^1, x^2,\pm\frac{\ell^2-\mathbf{x}\cdot \mathbf{x}}{2\ell}\right)~,
\label{coordstereoap}
\end{equation}
where $\mathbf{x} = (x^1,x^2) \in \mathbb{R}^2$ denotes a point on the two-sphere and the choice of $\pm$ in \eqref{coordstereoap} determines whether the origin corresponds to the south or north pole. The equator of the $S^2$ lies along $\mathbf{x}\cdot\mathbf{x}=\ell^2$ for either choice of sign. In these coordinates, our metric is: 
\begin{equation}
\dd s^2= \frac{4 \ell^4\dd\mathbf{x}\cdot \dd \mathbf{x}}{(\ell^2+ \mathbf{x}\cdot \mathbf{x})^2}\equiv \Omega(\mathbf{x})^2\dd\mathbf{x}\cdot \dd \mathbf{x}~,
\end{equation}
from which we can read off the conformal factor $\Omega(\mathbf{x})=\frac{2 \ell^2}{(\ell^2+ \mathbf{x}\cdot \mathbf{x})}$. In stereographic coordinates, the zweibeine are very easy to express: 
\begin{equation}\label{eq:zweibeinap}
e_\mu^a=\Omega(\mathbf{x})\,\delta^a_\mu~,\qquad\qquad e^\mu_a=\Omega(\mathbf{x})^{-1}\delta_a^\mu~.
\end{equation}

We will also need to define our Levi-Civita tensors. In Euclidean signature we will take: 
\begin{equation}\label{eq:levicivita}
\epsilon^{\mu\nu}=\frac{\tilde{\epsilon}^{\mu\nu}}{\sqrt{g}}~,\qquad\qquad \epsilon_{\mu\nu}=\sqrt{g}\,\tilde{\epsilon}_{\mu\nu}~,\qquad\qquad \tilde{\epsilon}^{12}=-\tilde{\epsilon}^{21}=\tilde{\epsilon}^{\vartheta\varphi}=-\tilde{\epsilon}^{\varphi\vartheta}=+1~.
\end{equation}
With these choices, we can verify that 
\begin{equation}
\gamma_\mu\gamma_*=-i\epsilon_{\mu\nu}\gamma^\nu~.\label{ap:2dcommute}
\end{equation}

Continuing to Lorentzian signature takes $u_{xy}^E\rightarrow u_{xy}^L$, where 
\begin{itemize}
    \item $u_{xy}^L<0$ for timelike-separated points,
    \item $0<u_{xy}^L<2$ for spacelike-separated points,
    \item $u_{xy}^L=0$ for null-separated points and $u_{xy}^L=2$ for antipodally separated points,
    \item $u_{xy}^L>2$ for spacelike spearated points which do not admit a geodesic connecting them.
\end{itemize}

\subsection{Coordinate invariant definition of Clifford algebra}\label{ap:cliffordJayewardena}

For those who wish to compare our formulas with those of \cite{Jayewardena:1988td}, it will be useful to provide a coordinate-independent prescription for defining the Clifford algebra elements. We let, $\vec{\mathbf{r}}\in \mathbb{R}^3$ be a set of embedding coordinates for the $S^2$ in $\mathbb{R}^3$ satisfying:
\begin{equation}
\vec{\mathbf{r}}\cdot \vec{\mathbf{r}}=\ell^2~.
\end{equation}
A few lines of algebra involving \eqref{eq:embmetric} are enough to show that the following matrices
\begin{equation}\label{ap:coordinvclifford}
\Gamma_\mu\equiv \frac{1}{\ell}\vec{\boldsymbol{\sigma}}\cdot\left(\vec{r}\times\partial_\mu\vec{r}\right)
\end{equation}
are a basis for the Clifford algebra, meaning they satisfy:
\begin{equation}
\left\{\Gamma_\mu,\Gamma_\nu\right\}=2g_{\mu\nu}\,\mathds{1}_{2\times 2}
\end{equation}
as in \eqref{eq:clifford}. Since this is an equivalent, albeit nonstandard, representation of the Clifford algebra which makes no reference to the frame fields $e^a_\mu$, we chose not use these matrices, breaking with the conventions in \cite{Jayewardena:1988td}. However, it must be the case that the two representations of the algebra $\Gamma_\mu$ and $\gamma_\mu$ are unitarily related to one another, meaning there exists some $u\in SU(2)$ such that: 
\begin{equation}\label{ap:gammarelation}
u\Gamma_\mu u^{-1}=\gamma_\mu~.
\end{equation}

Working in stereographic coordinates \eqref{coordstereoap} with frame fields given in \eqref{eq:zweibeinap}, we can provide such an $SU(2)$ matrix:
\begin{equation}\label{ap:umatrix}
u(\mathbf{x}) \equiv a_0(\mathbf{x})\left(\mathds{1}_{2\times 2}+i {\alpha}_i (\mathbf{x}) {\sigma}^i \right) 
\end{equation}
with 
\begin{equation}
a_0(\mathbf{x}) = \frac{\ell}{\sqrt{2(\ell^2+\mathbf{x}^2)}}~, \qquad \alpha_1 = \frac{x^1-x^2}{\ell}~, \qquad \alpha_2 = \frac{x^1 + x^2}{\ell}~, \qquad \alpha_3 = 1~.
\end{equation}
This matrix will make an appearance when we compute our fermionic two-point functions in the main text. 

\subsection{Embedding-space construction of the instanton configuration}\label{ap:monopole}
We will use the embedding formalism to construct the 1-instanton configurations of size $\ell$ on $S^2$ as in \eqref{eq:sizeoneinstanton}, in any coordinate system.  Letting, as before, $\vec{\mathbf{r}}\in \mathbb{R}^3$ be a vector that satisfies
\begin{equation}
\vec{\mathbf{r}}\cdot \vec{\mathbf{r}}=\ell^2~.
\end{equation}
Now define the two-component spinors $\zeta$ and $\chi$ (and their complex conjugates $\bar\zeta$ and $\bar\chi$) such that: 
\begin{equation}
\vec{\mathbf{r}}\cdot\vec{\boldsymbol{\sigma}}\,\zeta=\ell\, \zeta~,\qquad\qquad \vec{\mathbf{r}}\cdot\vec{\boldsymbol{\sigma}}\,\chi=-\ell\, \chi~,\qquad\qquad \bar{\zeta}\zeta=\bar{\chi}\chi=1~.
\end{equation}
With these definitions, we can show that: 
\begin{equation}
C_\mu={i}\bar{\zeta}\partial_\mu\zeta=-i\bar{\chi}\partial_\mu\chi~
\end{equation}
is a gauge-field configuration with winding number 1, and satisfies $\nabla_\mu F^{\mu\nu}=0$.

\subsection{Expansions on \texorpdfstring{$S^2$}{s2}}\label{ap:sphH}

Scalar functions $\Phi(\mathbf{x})$ defined on $S^2$ can be expanded in a complete basis of eigenfunctions of the two-sphere Laplacian as 
\begin{equation}\label{eq:bosonexpansion}
    \Phi(\mathbf{x}) = \sum_{L=0}^\infty \sum_{M=-L}^L c_{LM} Y_{LM}(\mathbf{x}) \, ,
\end{equation}
where the $Y_{LM}(\mathbf{x})$ are a basis of $\emph{real}$ spherical harmonics that satisfy the standard orthonormality conditions 
\begin{equation}\label{eq:YLM}
    -\nabla^2_{{S}^2} Y_{LM} = \frac{L(L+1)}{\ell^2} Y_{LM} \, , \qquad \int_{S^2} \textnormal{d}^2 x \sqrt{g}\, Y_{LM} Y_{L'M'} = \ell^2 \delta_{LL'} \delta_{MM'} \, ,
\end{equation}
as well as the completeness relation 
\begin{equation}
    \sum_{L=0}^\infty \sum_{M=-L}^L Y_{LM}(\mathbf{x}) Y_{LM}(\mathbf{y}) = \ell^2\frac{\delta(\mathbf{x} - \mathbf{y})}{\sqrt{g}} \, .
\end{equation}
Furthermore, we have the addition theorem 
\begin{equation}
    \sum_{M=-L}^L Y_{LM}(\mathbf{x}) Y_{LM}(\mathbf{y}) = \frac{2L+1}{4\pi} P_L\left( \cos \Theta_{xy} \right) \, .
\end{equation}

Similarly, fermionic fields $\Psi(\mathbf{x})$ can be expanded in a complete basis with Grassmann-valued coefficients 
\begin{equation} \label{fermExpgr}
    \Psi(\mathbf{x}) =  \sum_n c_n \psi_n(\mathbf{x}) \, , \qquad \bar{\Psi}(\mathbf{x}) =  \sum_n \psi_n^\dagger(\mathbf{x}) \bar{c}_n
\end{equation}
Recall that for Grassmann-valued integrals we have 
\begin{equation}
\begin{split}
    \int \dd c_n &= 0 \,, \qquad \int \dd c_n c_n = 1 \, , \\ 
    \int \dd \bar{c}_n &= 0 \,, \qquad \int \dd \bar{c}_n \bar{c}_n = 1 \, ,
\end{split}
\end{equation}
thus, in order to have a non-trivial result we need to saturate each integral with the corresponding Grassmann coefficient. The eigenfunctions of the Dirac operator on the sphere, $\psi_n(\mathbf{x}), \psi_n^\dagger(\mathbf{x})$ are known 
\begin{equation}\label{eq:spheredirac}
    \slashed{\nabla} \psi_n(\mathbf{x}) = \slashed{\nabla} \psi^{(s)}_{\pm, L M} (\mathbf{x}) = \pm i \frac{\left( L + 1 \right)}{\ell} \psi^{(s)}_{\pm, L M} (\mathbf{x}) \, ,\qquad s = \pm \, .
\end{equation}
where $L$ corresponds to a $SO(3)$ quantum number and $M$ to the $SO(2)$ and we have $L = 0, 1, \cdots$, $M=0,1,\cdots, L$. Working in the coordinate system \eqref{sphere}, the eigenfunctions take the form \cite{Camporesi:1995fb} 
\begin{equation}\label{k0eigenmodes}
    \psi^{(+)}_{\pm,L M}(\vartheta,\phi) = \frac{c_{L M}}{\sqrt{2}} e^{i(M+\frac12) \varphi} \begin{pmatrix} \Phi_{L M}(\vartheta) \\ \pm i \Psi_{L M}(\vartheta) 
    \end{pmatrix} \, , \qquad  \psi^{(-)}_{\pm,L M}(\mathbf{x}) = \frac{c_{L M}}{\sqrt{2}} e^{-i(M+\frac12) \varphi} \begin{pmatrix} \pm i \Psi_{L M}(\vartheta) \\ \Phi_{L M}(\vartheta)  \end{pmatrix} \, ,
\end{equation}
where 
\begin{equation}
    \begin{split}
        \Phi_{L M}(\vartheta) &= \cos^{M+1} \left(\frac{\vartheta}{2} + \frac{\pi}{4} \right) \sin^M\left(\frac{\vartheta}{2} + \frac{\pi}{4} \right) P_{L-M}^{(M,M+1)}(-\sin \vartheta) \, , \\
        \Psi_{L M}(\vartheta) &= \cos^M \left(\frac{\vartheta}{2} + \frac{\pi}{4} \right) \sin^{M+1} \left(\frac{\vartheta}{2} + \frac{\pi}{4} \right) P_{L-M}^{(M+1,M)}(-\sin\vartheta) = (-1)^{L-M} \Phi_{L M}\left(\frac{\pi}{2} - \vartheta\right) \, ,
    \end{split}
\end{equation}
where $P_{n}^{(a,b)}(x)$ are the real-valued Jacobi polynomials. The coefficients satisfy 
\begin{equation}
    c_{L M}^2 = \frac{(L+M+1)!(L-M)!}{2\pi L!^2} \, ,
\end{equation}
and ensure the following orthogonality conditions 
\begin{equation} \label{fermionorm}
    \int_{S^2} \dd^2 x \sqrt{g} \psi^{(s) \dagger}_{P L M}(\mathbf{x}) \psi^{(s')}_{P' L' M'} (\mathbf{x}) = \ell \delta_{L L'} \delta_{M M'} \delta_{P P'} \delta_{s s'} \, , \qquad P,P' = \pm  .
\end{equation}

\section{Path integral definitions and regularizations} \label{app:PIandreg}

\subsection{Bosonic path integrals}
 Writing down a path integral without specifying a regularization scheme is a meaningless endeavor. Since this paper mostly deals with Gaussian path integrals, we can be explicit and discuss a scheme that is particularly suited to our purposes. To demonstrate the issue (and set conventions), let us start with the simple example of a canonically normalized scalar, on $S^2$ and compute the path integral 
 \begin{equation}
Z_\phi=\int D\phi\, e^{-\frac{1}{2}\int_{S^2} \phi \left(-\nabla^2+m^2\right)\phi}~.
\end{equation}
Expanding the boson in a complete basis, as in \eqref{eq:bosonexpansion}, and taking
\begin{equation}
D\phi=\prod_{L=0}^{\infty}\prod_{M=-L}^L \dd c_{LM}
\end{equation}
we can write this as: 
\begin{align}
Z_\phi&=\int \prod_{L=0}^{\infty}\prod_{M=-L}^L \dd c_{LM} e^{-\frac{1}{2}\left(L(L+1)+m^2\ell^2\right)c_{LM}^2}=\prod_{L=0}^{\infty}\prod_{M=-L}^L\sqrt{\frac{2\pi}{L(L+1)+m^2\ell^2}}~, \\
&\text{`='}\text{det}^{-\frac{1}{2}}\left(-\nabla^2+m^2\right)\text{det}^{\frac{1}{2}}\left(\frac{2\pi}{\ell^2}\right)
\end{align}
which follows from \eqref{eq:YLM}. A naive computation of the top line would yield zero, stemming from the fact that we are taking the product over an infinite number of modes. In the second line we have replaced this product with a suggestive, albeit ill-defined, expression, given by the determinant of a local operator times a divergent constant; essentially, zero times infinity. It is therefore impossible to make sense of this expression without regularization. 

More generally, for any positive definite operator $-D^2$,\footnote{We denote the positive operator as $-D^2$ in analogy to the positive operator $-\nabla^2$.} for now devoid of zero-modes, it is standard to write down the following ill-defined equality:  
\begin{equation}\label{eq:bosonicgaussianpi}
Z_\phi=\int D\phi\, e^{-\frac{1}{2}\int \phi\left(- D^2\right)\phi} ~\text{`='}~ c_{D^2}\,\text{det}^{-\frac{1}{2}}\left(-D^2\right)~.
\end{equation}
This quantity is ill-defined for the usual reasons: we use `$=$' because the multiplicative constant $c_{D^2}$ is naively divergent while the determinant of $-D^2$ also diverges owing to $-D^2$ having an unbounded spectrum in the ultraviolet. The product of these two quantities needs to therefore be regularized. We start by taking the $\log$ of the above expression: 
\begin{equation}
\log Z_\phi ~\text{`='}~ \log(c_{D^2})-\frac{1}{2}\text{Tr}\log \left(-D^2\right)~, 
\end{equation}
but this clearly provides no additional prescription for dealing with the ambiguities, as it is the difference between two divergent quantities. We will deal with these issues by using \emph{heat kernel regularization} (see \cite{vassilevich} for a definitive reference or \cite{Denef:2009kn} for a nice exposition).  The idea is to first make use of the identity
\begin{equation}
    \log\left(\frac{\lambda}{\lambda_0}\right)=-\int_0^\infty \frac{d\tau}{\tau}\left(e^{-\lambda\tau}-e^{-\lambda_0\tau}\right)~,
\end{equation}
 to swap the order of the $\log$ and the trace. Our regulated quantity will then look as follows: 
\begin{equation}
\log Z_\phi^{f,\epsilon}=\int_0^\infty\frac{d\tau}{2\tau} f\left(\frac{\tau}{\epsilon^2}\right)\text{Tr}~e^{-\left(-D^2\right)\tau}~.
\end{equation}
In the above expression, we have introduced a \emph{regulator} $f(x)$, which must be taken to: a) go to 1 as $x\rightarrow \infty$, and b) go to zero sufficiently fast when $x\rightarrow 0$. Here $\epsilon^{-1}$ plays the role of a UV cutoff scale and smoothly suppresses the high-energy modes of $-D^2$. Naturally, any divergence is reintroduced if we take $\epsilon\rightarrow 0$. More specifically, the function $f(x)$ is a stand-in for a particular regularization scheme. For example, implementing a Pauli-Villars regulator in two-dimensions is equivalent to choosing
\begin{equation}
    f_{\rm PV}(x)=\left(1-e^{-x}\right)^k ~, \qquad k\geq \frac{3}{2}~.
\end{equation}
However, we will not choose to regulate in this way. Rather, we will use the \emph{Harish-Chandra} regulator introduced in \cite{Anninos:2020hfj}:
\begin{equation}
    f_{\rm HC}(x)=e^{-\frac{1}{4x}}~.
\end{equation}
We would like to emphasize that the choice of regulator will not affect any UV-finite answers, and ultimately, our choice of regulator is made for convenience. 

By way of notation we now write:
\begin{equation}\label{eq:heatkernelharishboson}
    \log {\det}_\epsilon\, \left(-D^2\right)\equiv -\int_0^\infty\frac{d\tau}{\tau} e^{-\frac{\epsilon^2}{4\tau}}\text{Tr}~e^{+D^2\tau}
\end{equation}
and when we wish to compute Gaussian bosonic path integrals, we will replace $Z_\phi$ as defined in \eqref{eq:bosonicgaussianpi} with: 
\begin{equation}
Z_\phi\rightarrow Z_\phi^\epsilon\equiv \exp \left[-\frac{1}{2} \log \text{det}_\epsilon\, \left(-D^2\right)\right]~.
\end{equation}
Note that in these expressions, the operator $D^2$ has units of $[\text{length}]^{-2}$, meaning $\epsilon$ has units of $[\text{length}]$. For bosons, we can define an emergent UV cutoff scale: 
\begin{equation}\label{eq:cutoffboson}
    \ell_{\rm UV}^\phi = \frac{\epsilon e^{\gamma}}{2} \, , \qquad \Lambda_{\rm UV}^\phi = \frac{1}{ \ell_{\rm UV}^\phi} \, ,
\end{equation}
associated with this regularization scheme where $\gamma\approx 0.5772$ is the Euler-Mascheroni constant. What this means in practice, is that, although the operator $-D^2$ whose determinant we are computing is dimensionful, the final regulated answer will be dimensionless, expressed in appropriate ratios with the emergent UV cutoff scale. 

\subsection{Fermionic path integrals}
Our procedure for regulating fermionic path integrals will follow the logic of the preceding section. However, in light of the fact that the differential operator of interest to us, \eqref{eq:nablaslashedk},  has $|k|$ zero-modes, we will add details in this section on how to treat the fermionic path integral in this situation. To avoid confusion, we will now refer to the differential operator appearing in our fermionic path integral as $\slashed{\mathcal{D}}$, and we will assume $\slashed{\mathcal{D}}$ has $n$ zero-modes.  An additional complication is that fermionic determinants also have a phase ambiguity (see \cite{Witten:2015aba} for a nice explanation of this fact), which we will discuss now. We start with the basic ambiguous definition mimicking \eqref{eq:bosonicgaussianpi}:
\begin{equation}\label{eq:fermiongaussian}
Z_{\psi}=\int D\overline{\psi}D\psi\, e^{-\int \overline{\psi} \slashed{\mathcal{D}}\psi} ~\text{`='}~ c_{\slashed{\mathcal{D}}^2}\det\left(\slashed{\mathcal{D}}\right)~.
\end{equation}

Besides needing to deal with the UV ambiguities and the aforementioned zero-modes in the above expression, the additional technical challenge for fermionic determinants is that the non-zero spectrum of $\slashed{\mathcal{D}}$ is often pure imaginary. This is expected, $\slashed{\mathcal{D}}$ is constructed as the `square root' of a bosonic operator with a negative-definite spectrum (for example, see \eqref{eq:spheredirac}). However, in certain cases, as will be the case in this paper, the non-zero spectrum of $\slashed{\mathcal{D}}$ comes in complex conjugate pairs: 
\begin{equation}
\slashed{\mathcal{D}}\psi_j^\pm=\pm i \lambda_j\psi_j^\pm~.
\end{equation}
When this happens, we can write the following formal expression \cite{Witten:2015aba}:
\begin{equation}
\text{det}' \slashed{\mathcal{D}}~\text{`='}~ \prod_j(+i\lambda_j)\prod_j(-i \lambda_j)=\prod_j\lambda_j^2~,
\end{equation}
where the prime indicates that we are taking the product over the non-zero-modes of $\slashed{\mathcal{D}}$.
Thus when the eigenvalues are paired up in this way, the phase ambiguity of this determinant can be regularized to zero, and we can write:
\begin{equation}\label{fermDetdef}
     \log \text{det}'_{\tilde{\epsilon}}\, \slashed{\mathcal{D}}={ \frac{1}{2}}\log \text{det}'_{\tilde{\epsilon}}\left(-\slashed{\mathcal{D}}^2\right)~,
\end{equation}
where the 1/2 accounts for halving the number of eigenvalues of $\slashed{\mathcal{D}}^2$ and the right hand side is regulated according to \eqref{eq:heatkernelharishboson}, but now $\tilde{\epsilon}$ is related to a different cutoff scale than for bosons (see equation \eqref{eq:cutofffermion} below).\footnote{Unless we have supersymmetry or some other organizing principle, we may assume a different cutoff scale for each field present in our model.} We provide this expression, because it is often easier to calculate the expression on the right hand side than to deal with the left hand side directly.

Now let us rewrite fermionic path integral following \eqref{fermDetdef}: 
\begin{align}\label{eq:fermionicgaussianpi}
Z_{\psi,s}=\int D\overline{\psi}D\psi\, e^{-\int \overline{\psi} \left(\slashed{\mathcal{D}}\pm\tfrac{s}{\ell}\right)\psi} &~\text{`='}~ c_{\slashed{\mathcal{D}}^2}\left[\det\left(\slashed{\mathcal{D}}+\tfrac{s}{\ell}\right)\det\left(-\slashed{\mathcal{D}}+\tfrac{s}{\ell}\right)\right]^{\frac{1}{2}}~,\nonumber\\
&~\text{`='}~ c_{\slashed{\mathcal{D}}^2}\text{det}^{\frac{1}{2}}\left(-\slashed{\mathcal{D}}^2+\tfrac{s^2}{\ell^2}\right)~,
\end{align}
where we have shifted the equality in \eqref{fermDetdef} by the small quantity $s^2/\ell^2$ because, otherwise, the above expression would formally vanish due to the existence of the zero-modes.\footnote{The radius of the sphere $\ell$ makes an appearance here simply to keep $s$ dimensionless.}  For any finite nonzero $s$, however, this expression is ambiguous and must be regulated. Proceeding as before, we replace: 
\begin{equation}
Z_{\psi,s}\rightarrow Z_{\psi,s}^{\tilde{\epsilon}}\equiv\exp\left[\frac{1}{2}\log\text{det}_{\tilde{\epsilon}}\left(-\slashed{\mathcal{D}}^2+\tfrac{s^2}{\ell^2}\right)\right]~, 
\end{equation}
with 
\begin{equation}
  \frac{1}{2}  \log \text{det}_{\tilde{\epsilon}}\,\left(-\slashed{\mathcal{D}}^2+\tfrac{s^2}{\ell^2}\right)= -\int_0^\infty\frac{d\tau}{2\tau} e^{-\frac{\tilde{\epsilon}^2}{4\tau}}\text{Tr}~e^{- \left(-\slashed{\mathcal{D}}^2+\tfrac{s^2}{\ell^2}\right)\tau}~.
\end{equation}
As was the case for bosons, we can define a an emergent UV cutoff scale for fermions: 
\begin{equation}\label{eq:cutofffermion}
    \ell_{\rm UV}^\psi = \frac{\tilde{\epsilon} e^{\gamma}}{2} \, , \qquad \Lambda_{\rm UV}^\psi = \frac{1}{ \ell_{\rm UV}^\psi} \, .
\end{equation}
We will eventually take $s\rightarrow0$, so let us separate out its contribution in this limit, which comes solely from the $n$ zero-modes: 
\begin{equation}
   \lim_{s\rightarrow0} \frac{1}{2}\log \text{det}_{\tilde{\epsilon}}\,\left(-\slashed{\mathcal{D}}^2+\tfrac{s^2}{\ell^2}\right)=n\log\left(\frac{s}{\ell \Lambda_{\rm UV}^\psi}\right) -\int_0^\infty\frac{d\tau}{2\tau} e^{-\frac{\tilde{\epsilon}^2}{4\tau}}\text{Tr}'~e^{+ \slashed{\mathcal{D}}^2\tau}+\mathcal{O}(s)
\end{equation}
where $\text{Tr}'$ means we sum over the nonzero-modes of the operator $\slashed{\mathcal{D}}$. By way of notation we now write:
\begin{equation}
    \log \text{det}_{\tilde{\epsilon}}'\, \left(-\slashed{\mathcal{D}}^2\right)\equiv -\int_0^\infty\frac{d\tau}{\tau} e^{-\frac{\tilde{\epsilon}}{4\tau}}\text{Tr}'~e^{+\slashed{\mathcal{D}}^2\tau}~. 
\end{equation}
Thus we are tempted to make the substitution (in the $s\rightarrow0$ limit):
\begin{equation}
    Z_{\psi,s}\rightarrow Z_{\psi,s}^{\tilde{\epsilon}}\equiv\left(\frac{s}{ \ell\Lambda_{\rm UV}^\psi}\right)^{n}\left[\text{det}_{\tilde{\epsilon}}'\, \left(-\slashed{\mathcal{D}}^2\right)\right]^{\frac{1}{2}}~,
\end{equation}
and note that this quantity vanishes in the strict $s\rightarrow0$ limit, as expected. This is almost our final expression. 

As an afterthought, the origin of the vanishing $s\rightarrow0$ limit is simply the unsaturated fermion zero-mode integrals. That is, if we expand our fermions as follows
\begin{equation}
    \psi(\mathbf{x}) =  \psi'(\mathbf{x})+\sum_{j=1}^n c_j \chi_j(\mathbf{x}) \, , \qquad 
    \overline{\psi}(\mathbf{x}) =  \overline{\psi}'(\mathbf{x})+\sum_{j=1}^n \bar{c}_j \bar{\chi}_j(\mathbf{x}) \, ,
\end{equation}
where $\chi_j$ and $\bar{\chi}_j$ for $j=1,\dots,n$ are a basis of the $n$-zero-modes and the $c_j$ and $\bar{c}_j$ are constant grassman numbers (and $\psi'(\mathbf{x})$ and $\overline{\psi}'(\mathbf{x})$ represent the part of the field configuration orthogonal to these zero-modes), this suggests the replacement: 
\begin{equation}\label{eq:regulatedfermipiwzeromodes}
Z_\psi\rightarrow Z_\psi^{\tilde{\epsilon}}=\left[\text{det}_{\tilde{\epsilon}}'\, \left(-\slashed{\mathcal{D}}^2\right)\right]^{\frac{1}{2}}\int \prod_{j=1}^n\frac{\dd \bar{c}_j\dd c_j}{\ell\Lambda_{\rm UV}^\psi}~.
\end{equation}
The above expression is useful, as, in the presence of sources, we will be able to saturate the zero-mode path integral and get a nonzero answer. Our regularization scheme has nevertheless provided us with a consistently chosen coefficient for these finitely-many integrals.

\section{Integrating out the fermions}\label{ap:integrateout}
Our goal is to compute the following fermionic path integral from the main text: 
\begin{equation}\label{fermionAPP}
   Z^\Psi_k\equiv\int D \bar{\Psi} D \Psi e^{-\int  \textnormal{d}^2 x \,\sqrt{g}\left[\bar{\Psi} \slashed{\nabla}_k \Psi-\bar\Psi e^{-\Phi\gamma_*}h^{-1}\eta-\bar\eta h e^{-\Phi\gamma_*}\Psi\right]}~,
\end{equation}
suitably regulated (see \cref{app:PIandreg}), and where $\slashed{\nabla}_k$ was defined in \eqref{eq:nablaslashedk}, but we repeat it here: 
\begin{equation}
    \slashed{\nabla}_k\equiv\frac{1}{\sqrt{g}}\begin{pmatrix}\Omega^{\frac{1-k}{2}} & 0 \\ 0 & \Omega^{\frac{1+k}{2}}\end{pmatrix}\begin{pmatrix}0 &\partial_{x^1}-i\partial_{x^2} \\ \partial_{x^1}+i\partial_{x^2} &0\end{pmatrix}\,\begin{pmatrix}\Omega^{\frac{1-k}{2}} & 0 \\ 0 & \Omega^{\frac{1+k}{2}}\end{pmatrix}~.
\end{equation}
The first step in computing the fermionic path integral is to find a Green's function for the operator $\slashed{\nabla}_k$, which satisfies
\begin{equation}\label{eq:greenfunctionk}
    \slashed{\nabla}_{k} S_k(\mathbf{x},\mathbf{y}) =\frac{\delta(\mathbf{x}-\mathbf{y})}{\sqrt{g}} \mathds{1}_{2\times2}~.
\end{equation}
At this stage the attentive reader might raise doubts about the existence of such an $S_k(\mathbf{x},\mathbf{y})$ given the presence of the $|k|$ zero-modes of $\slashed{\nabla}_k$ as previously reported in \eqref{eq:zeromodes}, which renders the differential operator non-invertible. While this is true, the existence of zero-modes simply means that such an $S_k$ cannot be unique, as we can add any number of zero-modes to it  and still satisfy the above equation. However, there is a natural choice for $S_k$, which we can deduce by the form of $\slashed{\nabla}_k$. In flat space ($\Omega=1$), we would like to solve the equation 
\begin{equation}\label{eq:Sflatdiffeq}
\boldsymbol{\sigma}\cdot\partial_{\mathbf{x}}\,S^{\rm flat}(\mathbf{x},\mathbf{y})=\delta(\mathbf{x}-\mathbf{y})\mathds{1}_{2\times2}
\end{equation}
whose solution we readily compute: 
\begin{equation}\label{sFlat}
S^{\rm flat}(\mathbf{x},\mathbf{y})  = \frac{1}{2\pi}  \frac{\boldsymbol{\sigma}\cdot (\mathbf{x}-\mathbf{y})}{|\mathbf{x}-\mathbf{y}|^2} ~.
\end{equation}
From this we deduce that 
\begin{equation} \label{SKResult}
S_k(\mathbf{x},\mathbf{y})  = \frac{1}{2\pi}  \begin{pmatrix}\Omega(\mathbf{x})^{-\frac{1-k}{2}} & 0 \\ 0 & \Omega(\mathbf{x})^{-\frac{1+k}{2}}\end{pmatrix}\frac{\boldsymbol{\sigma}\cdot (\mathbf{x}-\mathbf{y})}{|\mathbf{x}-\mathbf{y}|^2}  \begin{pmatrix}\Omega(\mathbf{y})^{-\frac{1-k}{2}} & 0 \\ 0 & \Omega(\mathbf{y})^{-\frac{1+k}{2}}\end{pmatrix}
\end{equation}
satisfies \eqref{eq:greenfunctionk}.

This path integral is Gaussian in the fermionic field variables, so we can integrate them out. To do so we will use the regularization procedure laid out in \cref{app:PIandreg}. Particularly, let us expand
\begin{equation}
    \Psi(\mathbf{x}) =  \Psi'(\mathbf{x})+\sum_{j=1}^{|k|} c_j \chi_j(\mathbf{x}) \, , \qquad 
    \bar{\Psi}(\mathbf{x}) =  \bar{\Psi}'(\mathbf{x})+\sum_{j=1}^{|k|} \bar{c}_j \bar{\chi}_j(\mathbf{x}) \, ,
\end{equation}
with $\chi_j$ the basis of $|k|$ zero-modes of the operator $\slashed{\nabla}_k$ given in \eqref{eq:zeromodes}, while $\Psi'(\mathbf{x})$ and $\bar{\Psi}'(\mathbf{x})$ are orthogonal to this zero-mode subspace. To simplify notation, let us define: 
\begin{equation}
      \hat{\eta}(\mathbf{x}) \equiv e^{-\Phi(\mathbf{x})\gamma_*} h^{-1}(\mathbf{x}) \eta(\mathbf{x}) \, , \qquad \hat{\overline{\eta}}(\mathbf{x}) = \bar{\eta}(\mathbf{x}) h(\mathbf{x}) e^{-\Phi(\mathbf{x}) \gamma_*} \, , 
\end{equation}
which means we now want to compute
\begin{align}
   Z^\Psi_k&=\int D \bar{\Psi} D \Psi e^{-\int  \textnormal{d}^2 x \,\sqrt{g}\left[\bar{\Psi} \slashed{\nabla}_k \Psi-\bar\Psi  \hat{\eta}-\hat{\overline{\eta}}\Psi\right]}~,\nonumber\\
   &\equiv Z_k^\chi \,Z_k^{\Psi'}~,
\end{align}
and we have split the path integral into a factor coming solely from the zero-modes and the rest of the field configurations. 
Following \eqref{eq:regulatedfermipiwzeromodes}, it is straightforward to show that the zero-mode path integral, suitably (re-)normalized, is given by: 
\begin{equation} \label{0ModePI}
    Z^\chi_k = \left( \frac{{1}}{\ell \Lambda_{\rm UV}^\Psi} \right)^{|k|} \prod_{j=1}^{|k|}\left(\hat{\overline{\eta}},\chi_j\right)\left(\bar{\chi}_j,\hat{\eta}\right)~,
\end{equation}
where we have introduced the compact notation: 
\begin{equation}
\left(f,g\right)\equiv \int\dd^2 x\sqrt{g}\,  f(\mathbf{x}) g(\mathbf{x})   \, .
\end{equation}
To obtain the contribution from the nonzero-modes, and defining $\overleftrightarrow{\slashed{\nabla}}_k\equiv\vec{\slashed{\nabla}}_k-\cev{\slashed{\nabla}}_k$, we must perform a few manipulations:
\begin{align}
   Z_k^{\Psi'}=&\int D \bar{\Psi}' D \Psi' e^{-\int  \textnormal{d}^2 x \,\sqrt{g}\left[\frac{1}{2}\bar{\Psi}' \overleftrightarrow{\slashed{\nabla}}_k \Psi'-\bar\Psi'  \hat{\eta}-\hat{\overline{\eta}}\Psi'\right]}~,\\
    =&~e^{ \int \dd^2 x \dd^2 y \sqrt{g_x} \sqrt{g_y} \hat{\bar{\eta}}(\mathbf{x}) S_k(\mathbf{x},\mathbf{y}) \hat{\eta}(\mathbf{y})}\nonumber\\
    &~~~~~~~~~~\times\int D \bar{\Psi}' D \Psi' e^{-\frac{1}{2}\int  \textnormal{d}^2 x \,\sqrt{g_x}\left[\left(\bar{\Psi}'(\mathbf{x})-\int\dd^2 y\sqrt{g_y}\,\hat{\overline{\eta}}(\mathbf{y})S_k(\mathbf{y},\mathbf{x}) \right) \overleftrightarrow{\slashed{\nabla}}_k \left(\Psi'(\mathbf{x})-\int\dd^2 y\sqrt{g_y}S_k(\mathbf{x},\mathbf{y}) \hat{\eta}(\mathbf{y})\right)\right]}~,
\end{align}
which we can summarize as
\begin{equation}
     Z_k^{\Psi'}\rightarrow Z_{k,\tilde{\epsilon}}^{\Psi'}=\left[{\det}'_{\tilde{\epsilon}} \left(-\slashed{\nabla}_k^2\right)\right]^{1/2}e^{ \int \dd^2 x\, \dd^2 y \sqrt{g_x} \sqrt{g_y} \,\hat{\overline{\eta}}(\mathbf{x}) S_k(\mathbf{x},\mathbf{y}) \hat{\eta}(\mathbf{y})}~.
\end{equation}
Putting everything together we have 
\begin{equation}\label{eq:fermipathintegralwdet}
  \boxed{  Z^\Psi_{k,\tilde{\epsilon}}= \left( \frac{1}{\ell \Lambda_{\rm UV}^\Psi} \right)^{|k|}\left[ \prod_{j=1}^{|k|}\left(\hat{\overline{\eta}},\chi_j\right)\left(\bar{\chi}_j,\hat{\eta}\right)\right]\left[{\det}'_{\tilde{\epsilon}} \left(-\slashed{\nabla}_k^2\right)\right]^{1/2}e^{ \int \dd^2 x\, \dd^2 y \sqrt{g_x} \sqrt{g_y} \,\hat{\overline{\eta}}(\mathbf{x}) S_k(\mathbf{x},\mathbf{y}) \hat{\eta}(\mathbf{y})} \, }~.
\end{equation}
In the next appendix, we explicitly compute the determinant appearing in the above expression.

\subsection{Heat kernel computation of fermionic determinants}\label{HK}
Having defined our regularization scheme, in this section we explicitly compute the regularized functional determinant $\left[{\det}'_{\tilde{\epsilon}} \left(-\slashed{\nabla}_k^2\right)\right]^{1/2}$, as well as the ratio $\left[\frac{{\det}_{\tilde{\epsilon}}'\left(-\slashed{\nabla}_k^2\right)}{{\det}_{\tilde{\epsilon}}\left(-\slashed{\nabla}_0^2\right)}\right]^{1/2}$ that will appear in the computation of fermionic $2|k|$-point functions. This computation appeared previously in \cite{Grewal:2021bsu}, which we repeat here.  We have:
\begin{equation}
    \frac{1}{2}\log {\det}'_{\tilde{\epsilon}}\left(-\slashed{\nabla}_k^2\right)= - \int_0^\infty \frac{\dd \tau}{2\tau} e^{-\frac{\tilde{\epsilon}^2}{4\tau}} \textnormal{Tr}' e^{+\tau \slashed{\nabla}_k^2} \, ,
\end{equation}
where we are using the heat kernel regulator discussed in appendix \ref{app:PIandreg}, and the trace $\text{Tr}'$ is understood to be taken over the entire set of non-zero eigenvalues of the corresponding operator $\slashed{\nabla}_k$. For the case of $k=0$ the result is known and can be found in \cite{Camporesi:1995fb}. For $k \neq 0$ one has the spectrum \cite{Jayewardena:1988td,Grewal:2021bsu} 
\begin{equation}\label{evalsk}
    \slashed{\nabla}_k \psi_L^\pm (\mathbf{x}) = \lambda_L^\pm \psi_L^\pm(\mathbf{x}) \, , \qquad \lambda_L^\pm = \pm\frac{ i}{\ell} \sqrt{(L+1)(L+1 + \lvert k \lvert)}\, , \quad L =0, 1, 2, \dots \, ,
\end{equation}
with degeneracies 
\begin{equation}
    \textnormal{deg}\lambda_L^+ =\textnormal{deg}\lambda_L^- = 2(L+1) + \lvert k \lvert  \, .
\end{equation}
Now we express the heat kernel as 
\begin{equation}\label{newdet}
   \frac{1}{2}\log {\det}'_{\tilde{\epsilon}}\left(-\slashed{\nabla}_k^2\right)= - \int_0^\infty \frac{d\tau}{\tau}e^{-\frac{\tilde{\epsilon}^2}{4\tau}}  \sum_{L=0}^\infty (2L+2+|k|) e^{-\tau \frac{(L+1)(L+1 + | k | )  }{\ell^2} }~,
\end{equation}
where $\tilde{\epsilon}$ has units of length and is related to an emergent UV cutoff scale as in \eqref{eq:cutofffermion}. At the outset, we are stuck computing the infinite sum in the above expression. To proceed, we exploit the techniques developed in section 3 of \cite{Anninos:2020hfj} and start by completing the square 
\begin{equation}
(L+1)(L+1 + | k | )  = \left(L+1+\frac{|k|}{2}\right)^2-\frac{k^2}{4}~,
\end{equation}
and rearrange the sum inside the integrand as follows
\begin{equation}
    \sum_{n=1}^\infty (2n+|k|) e^{-\frac{\tau}{\ell^2}\left( n+\frac{|k|}{2} \right)^2 } = \int_{\mathbb{R}+i\delta} \dd u \frac{e^{-\frac{u^2 \ell^2}{4\tau}}}{\sqrt{4\pi \tau \ell^{-2}}} \sum_{n=1}^\infty (2n+|k|)e^{iu \left(n+\frac{|k|}{2}\right)} \, ,
\end{equation}
where the contour is deformed slightly away from the real axis with a small and positive $\delta > 0$ in order to render the sum on the right-hand side convergent. Additionally, we have changed the summation variable to $n=L+1$, whose range is over $\mathbb{Z}^+$. Performing the sum, we find
\begin{align}\label{sum}
\sum_{n=1}^\infty (2n+|k|)e^{iu \left(n+\frac{|k|}{2}\right)} &=-2i\partial_u\,\sum_{n=1}^\infty e^{iu \left(n+\frac{|k|}{2}\right)}~,\nonumber\\
&=\partial_u\frac{e^{i(|k|+1)\frac{u}{2}}}{\sin\frac{u}{2}}~,
\end{align}
which diverges quadratically as $u\to 2\pi \mathbb{Z}$, but is otherwise finite. Since we are integrating along the real axis with an imaginary shift $\delta$, our contour does not cross these poles. Going back to (\ref{newdet}) and substituting (\ref{sum}) for the sum, we arrive at the expression
\begin{align}
 \frac{1}{2}\log {\det}'_{\tilde{\epsilon}}\left(-\slashed{\nabla}_k^2\right) &= - \int_0^\infty \frac{\dd \tau}{\tau}\int_{\mathbb{R}+i\delta} \dd u\,\partial_u \left(\frac{e^{i(|k|+1)\frac{u}{2}}}{\sin\frac{u}{2}} \right)\frac{1}{ \sqrt{4\pi \tau \ell^{-2}}} e^{-\frac{\tilde{\epsilon}^2+u^2\ell^2}{4\tau}+\frac{k^2\tau}{4\ell^2} }   \, ~,\\
 &= - \int_0^\infty \frac{\dd \tau}{\tau}\int_{\mathbb{R}+i\delta} \dd u\, \frac{u }{\sin\frac{u}{2}}e^{i(|k|+1)\frac{u}{2}} \left(\frac{\ell^2}{\tau}\right)^{3/2}\frac{ e^{-\frac{\tilde{\epsilon}^2+u^2\ell^2}{4\tau}+\frac{k^2\tau}{4\ell^2} }}{4 \sqrt{\pi}}    \, ~,
\end{align}
where we have integrated by parts in going from the first to the second line. We would now like to continue by performing the $\tau$ integral before the $u$ integral, but note that at fixed $u$, the integrand diverges at large $\tau$. We can deal with this by yet another Hubbard-Stratonovich trick, rewriting the above expression as:
\begin{equation}
\frac{1}{2}\log {\det}'_{\tilde{\epsilon}}\left(-\slashed{\nabla}_k^2\right)=- \int_0^\infty \frac{\dd \tau}{\tau}\int_{\mathbb{R}+i\delta} \dd u\, \frac{u }{\sin\frac{u}{2}}e^{i(|k|+1)\frac{u}{2}} \int_{-\infty}^\infty \dd v\,\frac{\ell^4}{8\pi\tau^2}e^{-\frac{\tilde{\epsilon}^2+\left(u^2+v^2\right)\ell^2}{4\tau}+\frac{|k| v}{2} }~.
\end{equation}
We now swap the order of the $\tau$ and $v$ integration, as the $\tau$ integral has become absolutely convergent for any fixed $u$ and $v$: 
\begin{equation}
\frac{1}{2}\log {\det}'_{\tilde{\epsilon}}\left(-\slashed{\nabla}_k^2\right)=- \int_{\mathbb{R}+i\delta} \dd u\int_{-\infty}^\infty \dd v\, \frac{2u }{\pi\sin\frac{u}{2}}\frac{e^{i(|k|+1)\frac{u}{2}+ |k|\frac{v}{2}}}{\left(u^2+v^2+\frac{\tilde{\epsilon}^2}{\ell^2}\right)^2}~.
\end{equation}
The next step is to note that we can swap the order of the $u$ and $v$ integrals, and calculate the $u$-integral using the residue theorem. Since the $u$ contour misses the poles along the real axis, we can simply close the contour in the upper half plane and pick up the residue at $u=+i\sqrt{v^2+\left(\frac{\tilde{\epsilon}}{\ell}\right)^2}$~. Note that we will always pick up this pole in the upper half-plane, provided $0 < \delta < \frac{\tilde{\epsilon}}{\ell}$. We finally are left to compute the following: 
\begin{equation}
\frac{1}{2}\log {\det}'_{\tilde{\epsilon}}\left(-\slashed{\nabla}_k^2\right)=\int_{-\infty}^\infty \dd v\,\frac{e^{ |k| \frac{v}{2}}}{v}\partial_v\left(\frac{e^{-\frac{|k|+1}{2}\sqrt{v^2+\frac{\tilde{\epsilon}^2}{\ell^2}}}}{\sinh\left(\tfrac{1}{2}\sqrt{v^2+\frac{\tilde{\epsilon}^2}{\ell^2}}\right)}\right)~.
\end{equation}

To connect with the literature on Harish-Chandra characters, it will be useful to write this expression more similarly to \cite{Anninos:2020hfj}. To do this, we split the above integral over positive and negative $v$ and write it as a single integral over $v>0$: 
\begin{equation}\label{detknzv}
   \frac{1}{2}\log {\det}'_{\tilde{\epsilon}}\left(-\slashed{\nabla}_k^2\right) = - \int_{0}^\infty \frac{\dd v}{\sqrt{v^2 + \frac{\tilde{\epsilon}^2}{ \ell^{2}}}} \frac{4 + 2|k|\left(1-e^{-\sqrt{v^2 + \frac{\tilde{\epsilon}^2}{ \ell^{2}}}}\right)}{\left(1-e^{-\sqrt{v^2 + \frac{\tilde{\epsilon}^2}{ \ell^{2}}}}\right)^2} e^{-\left(\frac{|k|}{2}+1\right)\sqrt{v^2 + \frac{\tilde{\epsilon}^2}{ \ell^{2}}}}\cosh\left(|k|\frac{v}{2}\right) \, .
\end{equation}
Finally we make the substitution $v\rightarrow\sqrt{t^2-\frac{\tilde{\epsilon}^2}{{\ell^2}}}$ with $t>\frac{{\tilde{\epsilon}}}{{\ell}}$, yielding: 
\begin{equation}\label{detknz}
   \frac{1}{2}\log {\det}'_{\tilde{\epsilon}}\left(-\slashed{\nabla}_k^2\right) = - \int_{\frac{{\tilde{\epsilon}}}{{\ell}}}^\infty \frac{\dd t}{\sqrt{t^2 - \frac{\tilde{\epsilon}^2}{ \ell^{2}}}} \frac{4 + 2|k|\left(1-e^{-t}\right)}{\left(1-e^{-t}\right)^2} e^{-(|k|+2)\frac{t}{2}}\cosh\left(\frac{|k|}{2}\sqrt{t^2 - \frac{\tilde{\epsilon}^2}{ \ell^{2}}}\right) \, .
\end{equation}
which can be compared with various character expressions from \cite{Anninos:2020hfj,Grewal:2021bsu}. For example, Upon setting $k=0$ and formally\footnote{What we mean by `formally' is that we set $\tilde{\epsilon} = 0$ to yield a simpler looking expression. Whenever we evaluate the integral, we must reinstate $\tilde{\epsilon}$ as originally defined in our heat kernel procedure. Though we will generally take $\tilde{\epsilon}/\ell$ to be small and positive, it is understood to be finite.} setting $\tilde{\epsilon} = 0$ we find
\begin{equation}\label{detkz}
    \frac{1}{2}\log {\det}_{\tilde{\epsilon}=0}\left(-\slashed{\nabla}_{k=0}^2\right) =  - \int_0^\infty \frac{\dd t}{t} \frac{4 e^{-t}}{(1-e^{-t})^2}  \, ,
\end{equation}
which one can be compare with the two-dimensional version of expression (3.13) in \cite{Anninos:2020hfj} for the massless Dirac fermion. 

We now have a finite expression \eqref{detknzv} for the regularized determinant in any $k$-instanton sector. At present, we would like to extract the functional form of this integral in the small $\tilde{\epsilon}$ limit. This is subtle, and to do this carefully, we proceed in a slightly ad-hoc fashion. Let us fix $k>0$ without loss of generality. By comparing with the numerical evaluation of  \eqref{detknzv}, we note that three derivatives in $k$ can be computed accurately in this limit by setting $\tilde{\epsilon}=0$ inside the integrand, yielding:\footnote{Computing $\partial_k^2  \log {\det}'_{\tilde{\epsilon}}\left(-\slashed{\nabla}_k^2\right)$ in a similar fashion yields a finite result, but differs from the finite-$\tilde{\epsilon}$ numerical evaluation of the integral \eqref{detknzv} by order-1 shifts. Setting $\tilde{\epsilon}$ to zero in the integrand with  a single derivative in $k$ results in a divergence.  }
\begin{equation}
\lim_{\tilde{\epsilon}\rightarrow 0}\partial_k^3  \left[\frac{1}{2}\log {\det}'_{\tilde{\epsilon}}\left(-\slashed{\nabla}_k^2\right) \right] =  \psi ^{(1)}(k+1)+k \psi ^{(2)}(k+1)~,
\end{equation}
where $\psi^{(m)}(x) = \left(\frac{d}{d x}\right)^{m+1} \log \Gamma(x)$ is the Polygamma function.\footnote{For negative orders this is defined by \begin{equation*}\psi^{(-m)}(x) =\frac{1}{(m-2)!}\int_0^x(x-t)^{m-2}\log\Gamma(t)\dd t~.\end{equation*}} To continue, we must take three antiderivatives of the above expression, while fixing integration constants such that they match the numerical evalutation of derivatives with respect to $k$ of \eqref{detknzv} at small but finite $\tilde{\epsilon}$. Proceeding in this fashion, we find: 
\begin{multline}
\lim_{\tilde{\epsilon}\rightarrow 0}  \left[\frac{1}{2}\log {\det}'_{\tilde{\epsilon}}\left(-\slashed{\nabla}_k^2\right) \right] = \log\mathcal{N}^\Psi_{\tilde{\epsilon}}+|k|\log\left(\ell\Lambda_{\rm UV}^\Psi\frac{|k|!}{k^2}\right)\\-\frac{k^2}{2}+|k|\left(1+\log(2\pi)\right)-2\psi^{(-2)}(|k|)~,\label{eq:fermialmostthere}
\end{multline}
 where $\Lambda_{\rm UV}^\Psi$ is related to $\tilde{\epsilon}$ through \eqref{eq:cutofffermion}. The quantity $\mathcal{N}^\Psi_{\tilde{\epsilon}}$ is an overall $k$-independent normalization constant, which we give here for completeness:
\begin{equation}\label{eq:normalizationfermdet}
\log\mathcal{N}^\Psi_{\tilde{\epsilon}}=-\frac{4\ell^2}{\tilde{\epsilon}^2}-\frac{1}{3}\log\left(\frac{\tilde{\epsilon}}{\ell}\right)+0.700335 +\mathcal{O}\left(\frac{\tilde{\epsilon}}{\ell}\right)~.
\end{equation}
We obtain this expression by numerically evaluating \eqref{detknzv} and curve fitting. 

We recognize in \eqref{eq:fermialmostthere} the expression for the hyperfactorial function $K(z)$ \cite{wiki:K-function}, where for natural numbers $n$: 
\begin{equation}
 K(n)=\frac{[(n-1)!]^{n-1}}{G(n)}\equiv\exp\left[{\psi^{(-2)}(n)+\frac{n^2-n}{2}-\frac{n}{2}\log(2\pi)}\right]~,
\end{equation}
and where $G(n)$ is the Barnes function. Using this we can write: 
\begin{equation}
\lim_{\tilde{\epsilon}\rightarrow 0}  \left[\frac{1}{2}\log {\det}'_{\tilde{\epsilon}}\left(-\slashed{\nabla}_k^2\right) \right] = \log\mathcal{N}^\Psi_{\tilde{\epsilon}}+\frac{k^2}{2}+|k|\log\left({\ell\Lambda_{\rm UV}^\Psi}\right)-\log\left({|k|!}\right)-\sum_{l=1}^{|k|}\log\left(l^{2l-|k|-1}\right)~.
\end{equation}
This allows us to write: 
\begin{align} \label{ratioFdet}
  \left[\frac{{\det}_{\tilde{\epsilon}}'\left(-\slashed{\nabla}_k^2\right)}{{\det}_{\tilde{\epsilon}}\left(-\slashed{\nabla}_0^2\right)}\right]^{1/2} &= \left(\ell\Lambda_{\rm UV}^\Psi\right)^{|k|} \left(\frac{|k|!}{k^2}\right)^{|k|}e^ {-\frac{k^2}{2}+|k|\left(1+\log(2\pi)\right)-2\psi^{(-2)}(|k|)}  \, ,\\
  &=\left(\ell\Lambda_{\rm UV}^\Psi\right)^{|k|}\frac{e^{\frac{k^2}{2}}}{ |k|! \prod_{l=1}^{|k|} l^{2l-|k|-1}} ~.
\end{align}
Remarkably, the power of the cutoff in the above expression will precisely cancel against the zero-mode measure. Combining this with \eqref{eq:fermipathintegralwdet} we find: 
\begin{equation}\label{eq:fermipathintegralfinal}
  \boxed{  Z^\Psi_{k,\tilde{\epsilon}}= \frac{\mathcal{N}^\Psi_{\tilde{\epsilon}}\,e^{\frac{k^2}{2}}}{ |k|! \prod_{l=1}^{|k|} l^{2l-|k|-1}}\left[ \prod_{j=1}^{|k|}\left(\hat{\overline{\eta}},\chi_j\right)\left(\bar{\chi}_j,\hat{\eta}\right)\right]e^{ \int \dd^2 x\, \dd^2 y \sqrt{g_x} \sqrt{g_y} \,\hat{\overline{\eta}}(\mathbf{x}) S_k(\mathbf{x},\mathbf{y}) \hat{\eta}(\mathbf{y})} \, }~.
\end{equation}

\section{Two-sphere partition function}\label{S2app}

In this appendix, we compute the simplest calculable of the Schwinger model on a two-sphere---the two-sphere partition function $Z_{S^2}$. The sphere partition function only receives contributions from the $k=0$ topological gauge field theoretic sector. As such, it is a relatively straightforward Gaussian path integral. Let us start from (\ref{exactSpf}):
\begin{equation} \label{zs2}
Z_{S^2} =  \int \frac{D \Phi D h}{\textnormal{vol} \, \mathcal{G}} J_{\Phi,h}\, e^{ - \frac{1}{2 q^2} \int \dd^2 x \sqrt{g} \left[ \Phi \nabla^2\left( \nabla^2 - \frac{q^2}{\pi}  \right) \Phi\right]} Z^\Psi_{k=0,\tilde{\epsilon}} ~.
\end{equation}
The Jacobian measure factors \cite{Klebanov:2011td,Giombi:2015haa} coming from the decomposition in (\ref{Ak}) is given, formally,  by
\begin{equation}\label{jac}
J_{\Phi,h} = {\left|{\det}' \partial_\mu | \times | {\det}' \epsilon_{\mu\nu} \partial^\nu \right|} \equiv {\det}{}' \left(-\nabla^2\right)~,
\end{equation}
where the prime indicates we are not including the constant mode, and we have used the differential form identity $\star \, \dd \star \dd = -\nabla^2$. The reason for not including the constant mode in (\ref{jac}) is that the constant configurations of $h$ and $\Phi$ do not contribute to the physical field $A_\mu$ in its decomposition (\ref{Ak}). We can alternatively interpret $J_{\Phi,h}$ as the determinant coming from path integrating over the Fadeev-Popov ghosts upon fixing the Lorenz gauge. The path integral over $h$ is canceled almost entirely by the volume $\text{vol }\mathcal{G}$ of the $U(1)$ gauge group. The remaining part, as discussed in detail in section 5 of \cite{Anninos:2020hfj}, is given by  the volume of the constant part of the $U(1)$ group, $\text{vol}\,U(1) = 2\pi$, such that
\begin{equation}
\frac{1}{\text{vol} \, \mathcal{G}}  \int  D h = \frac{1}{2\pi}~.
\end{equation}
The heat-kernel regularized fermion partition function $Z^\Psi_{k=0,\tilde{\epsilon}}$ is given by (\ref{detkz}) or, more specifically, in our scheme, by $\mathcal{N}_{\tilde{\epsilon}}^\Psi$ given in \eqref{eq:normalizationfermdet}. 

Since we are excluding zero-modes in our determinants, we must discuss one more subtlety. Consider the general Gaussian path integral from \eqref{eq:bosonicgaussianpi}: 
\begin{equation}
\int D\phi\, e^{-\frac{1}{2}\int \phi\left(- D^2\right)\phi} ~\text{`='}~ c_{D^2}\,\text{det}^{-\frac{1}{2}}\left(-D^2\right)~, 
\end{equation}
where $-D^2$ should be thought of as any positive operator. Under the field redefinition $\phi(\mathbf{x})\rightarrow f(\mathbf{x})/w$, the answer should remain unchanged
\begin{equation}
\int Df\, e^{-\frac{1}{2w^2}\int f\left(- D^2\right)f} ~\text{`='}~ c_{\frac{D^2}{w^2}}\,\text{det}^{-\frac{1}{2}}\left(-\frac{D^2}{w^2}\right)=c_{D^2}\,\text{det}^{-\frac{1}{2}}\left(-D^2\right)~, 
\end{equation}
We can understand this equality if we expand the field $\phi$ in a complete set of modes $\phi(\mathbf{x})=\sum_ic_i\phi_i(\mathbf{x})$; then the measure transforms as
\begin{equation}
D\phi=\prod_i \dd c_i\rightarrow \prod_i \frac{\dd c_i}{w}=D f~,
\end{equation}
which soaks up the overall $w$ dependence, mode by mode. The subtlety now comes in if we have $n$ zero-modes and want to compute the determinant with zero-modes excluded. We first give the result before arguing for it: 
\begin{equation}\label{ap:zeromoderemovedextrafactor}
 c_{\frac{D^2}{w^2}}\,{\det}'^{-\frac{1}{2}}\left(-\frac{D^2}{w^2}\right)={w^n}c_{D^2}\,{\det}'^{-\frac{1}{2}}\left(-D^2\right)~.
\end{equation}
The extra factor of $w^{n}$ can be thought of as coming from $n$ uncancelled powers of $w$ which are there, roughly speaking, because the primed determinant comes from a path integral with $n$ missing measure factors---because we are not integrating over these $n$ zero-modes. This type of logic coincides with many alternative treatments \cite{Witten:1995gf, Giombi:2015haa,Law:2020cpj,Anninos:2020hfj} (see also \cite{Chandrasekaran:2022cip} above equation (4.22) for a general statement along these lines).

Integrating out $h$, $\Phi$, and $\Psi$ in the path integral (\ref{zs2}), one finds the following formal expression
\begin{equation} \label{ap:bosonoptiona}
Z_{S^2} ~\text{`='}~ \frac{q\ell}{2\pi}\mathcal{N}_{\tilde{\epsilon}}^\Psi\frac{{\det}' \left(-\nabla^2\right)}{\left[ {{{\det}' \,\left(-\nabla^2\left(-\nabla^2 + \frac{q^2 }{\pi}  \right)  \right)}} \right]^{1/2}} ~,
\end{equation}
where the $q\ell$ dependent prefactor follows from the logic leading to \eqref{ap:zeromoderemovedextrafactor}, noting that we are not path integrating over the unique zero-mode of $\Phi$. The factor of $\ell$ specifically appears in the above formula to soak up units. 
Naturally, given our discussion in \cref{app:PIandreg}, this quantity is meaningless without regularization.  However, before getting to it, let us massage this formal expression using known properties of finite determinants. Specifically, let us use $\det(AB)=\det(A)\det(B)$ in the above equation: 
\begin{equation}\label{ap:bosonoptionb}
Z_{S^2} ~\text{`='}~ \frac{q\ell}{2\pi}\mathcal{N}_{\tilde{\epsilon}}^\Psi\frac{\left[{\det}' \left(-\nabla^2\right)\right]^{1/2}}{\left[ {\det}'\left(-\nabla^2 +\frac{q^2 }{\pi}  \right)   \right]^{1/2}} ~,
\end{equation}
Regularizing \eqref{ap:bosonoptiona} versus \eqref{ap:bosonoptionb} using the methods provided in \cref{app:PIandreg} will lead to different results. But the two results must agree in the small $\epsilon$ limit up to an overall normalization of the path integral, which we already know is ambiguous. This ambiguity stems from the existence of a local counterterm $\int d^2 x\sqrt{g} R$ (proportional to the Euler character) which one can always add to the action in curved space. Let us state again for clarity that the overal normalization of \eqref{ap:bosonoptionb} will carry no physical information. Hence, in addition to our choice of scheme, we choose to define our regularized sphere path integral as follows: 
\begin{equation}\label{partition}
Z_{S^2}^{\epsilon\tilde{\epsilon}} = \frac{q\ell}{2\pi}\mathcal{N}_{\tilde{\epsilon}}^\Psi\frac{\left[{\det}_\epsilon' \left(-\nabla^2\right)\right]^{1/2}}{\left[ {\det}_\epsilon'\left(-\nabla^2 +\frac{q^2 }{\pi}  \right)   \right]^{1/2}} ~,
\end{equation}

Let us now take the logarithm of this expression
\begin{equation} \label{eq:logpart}
\log Z_{S^2}^{\epsilon\tilde{\epsilon}} \equiv \log\mathcal{N}_{\tilde{\epsilon}}^\Psi+\log\frac{ q\ell}{2\pi}+\frac{1}{2}\log{\det}_\epsilon' \left(-\nabla^2\right)-\frac{1}{2}\log{{{\det}'_{\epsilon} \,\left(-\nabla^2 + \frac{q^2 }{\pi}   \right)}} ~.
\end{equation}

In the $q\ell \to 0$ limit of vanishing coupling, the partition function $Z_{S^2}^{\epsilon\tilde{\epsilon}}$ becomes a product of a massless Dirac fermion partition function with that of a free $U(1)$ gauge theory. As a function $q\ell$, the partition function \eqref{partition} encodes the full quantum loop expansion of the sphere path integral. 

To compute this function, we start with the following term:
\begin{equation}
   -\frac{1}{2}\log {\det}'_{{\epsilon}}\left(-{\nabla}^2+ \frac{q^2}{\pi }\right)=  \int_0^\infty \frac{d\tau}{2\tau}e^{-\frac{{\epsilon}^2}{4\tau}}  \sum_{L=1}^\infty (2L+1) e^{-\tau\left[\frac{L(L+1)  }{\ell^2}+\frac{q^2}{\pi}\right]}~,
\end{equation}
but we are immediately thwarted by the sum in its current form. Proceeding exactly as we did in \cref{HK}, we first write this as
\begin{align}
 -\frac{1}{2}\log {\det}'_{{\epsilon}}\left(-{\nabla}^2+ \frac{q^2}{\pi }\right)&=\int_0^\infty \frac{d\tau}{2\tau}e^{-\frac{{\epsilon}^2}{4\tau}+\left(\frac{1}{4\ell^2}-\frac{q^2}{\pi}\right)\tau} \int_{\mathbb{R}+i\delta} \dd u \frac{e^{-\frac{u^2 \ell^2}{4\tau}}}{\sqrt{4\pi \tau \ell^{-2}}} \sum_{L=1}^\infty (2L+1)e^{iu \left(L+\frac{1}{2}\right)}\nonumber\\
 &=\int_0^\infty \frac{d\tau}{2\tau}e^{-\frac{{\epsilon}^2}{4\tau}+\left(\frac{1}{4\ell^2}-\frac{q^2}{\pi}\right)\tau} \int_{\mathbb{R}+i\delta} \dd u \frac{e^{-\frac{u^2 \ell^2}{4\tau}}}{\sqrt{4\pi \tau \ell^{-2}}} \partial_u \left(\frac{e^{i u}}{\sin\frac{u}{2}}\right)~.
\end{align}
where the $u$ contour is chosen to ensure convergence of the sum.\footnote{This is basically an $i\epsilon$ prescription, but we have already used the symbol $\epsilon$ to denote our regulator.} 
We now want to swap the order of the $\tau$ and $u$ integrals, but the $\tau$ integral is not absolutely convergent for any $u$ if $\frac{q^2\ell^2}{\pi}<\frac{1}{4}$. Hence, as in \cref{HK}, we perform a second Hubbard-Stratonovitch trick, yielding: 
\begin{equation}
 -\frac{1}{2}\log {\det}'_{{\epsilon}}\left(-{\nabla}^2+ \frac{q^2}{\pi }\right)=-\int_0^\infty \frac{d\tau}{2\tau} \int_{\mathbb{R}+i\delta} \dd u\int_{-\infty}^\infty\dd v \,\frac{e^{i u}}{\sin\frac{u}{2}}\,\partial_u\left(\frac{e^{-\frac{\epsilon^2+\left(u^2+v^2\right) \ell^2}{4\tau}+i\,\nu\,v}}{4\pi \tau\ell^{-2} }\right) ~.
\end{equation}
where we have defined $\nu \equiv \tfrac{1}{2} \sqrt{ \tfrac{4q^2 \ell^2}{\pi}-1}$ and have integrated by parts. We can now evaluate the above expression by performing the $\tau$-integral first, then by  closing the $u$ contour in the upper-half plane and using the residue theorem, leading to: 
\begin{equation}
  -\frac{1}{2}\log {\det}'_{{\epsilon}}\left(-{\nabla}^2+ \frac{q^2}{\pi }\right)=-\int_{-\infty}^\infty \dd v\,\frac{e^{ i\,\nu\,v}}{2v}\partial_v\left(\frac{e^{-\sqrt{v^2+\frac{{\epsilon}^2}{\ell^2}}}}{\sinh\left(\tfrac{1}{2}\sqrt{v^2+\frac{{\epsilon}^2}{\ell^2}}\right)}\right)~.
\end{equation}
Again it will be useful to write this expression more similarly to \cite{Anninos:2020hfj} by splitting the above integral over positive and negative $v$ and writing it as a single integral over $v>0$: 
\begin{equation}
    -\frac{1}{2}\log {\det}'_{{\epsilon}}\left(-{\nabla}^2+ \frac{q^2}{\pi }\right) = \int_{0}^\infty \frac{\dd v}{\sqrt{v^2 + \frac{{\epsilon}^2}{ \ell^{2}}}}\left( \frac{1+e^{-\sqrt{v^2 + \frac{{\epsilon}^2}{ \ell^{2}}}}}{\left(1-e^{-\sqrt{v^2 + \frac{{\epsilon}^2}{ \ell^{2}}}}\right)^2}-1\right) e^{-\frac{1}{2}\sqrt{v^2 + \frac{{\epsilon}^2}{ \ell^{2}}}}\cos\left(\nu \, v\right) \, .
\end{equation}
Finally, substituting $v\rightarrow\sqrt{t^2-\frac{{\epsilon}^2}{{\ell^2}}}$ with $t>\frac{{{\epsilon}}}{{\ell}}$, we find: 
\begin{multline}
     -\frac{1}{2} \log {{{{\det}'_\epsilon} \left( -\nabla^{2} + \frac{q^2}{\pi }   \right)}}  = \\\int_{\frac{\epsilon}{\ell}}^\infty \frac{\dd t}{2 \sqrt{t^2 - \frac{\epsilon^2}{\ell^2}}} \left[ \frac{1+e^{-t}}{(1-e^{-t})^2}  -1 \right]\left( e^{-\frac{t}{2}-i \nu \sqrt{t^2 -\frac{\epsilon^2}{\ell^2}}}+ e^{-\frac{t}{2}+ i \nu \sqrt{t^2-\frac{\epsilon^2}{\ell^2}}} \right)~.
\end{multline}
If we formally set $\epsilon = 0$ in the above expression, and defining $\Delta \equiv \frac{1}{2}+i \nu$, $\bar{\Delta} \equiv 1- \Delta$,  we find: 
\begin{equation} \label{detq2}
-\frac{1}{2} \log {{{{\det}'_\epsilon} \left( -\nabla^{2} + \frac{q^2}{\pi }   \right)}}  = \int_0^\infty \frac{dt}{2t} \left[ \frac{1+e^{-t}}{1-e^{-t}} \frac{e^{-\Delta t }+ e^{-\bar{\Delta} t  }}{1-e^{-t}} - \left( e^{-\Delta t }+ e^{-\bar{\Delta} t} \right)  \right] ~, 
\end{equation}
which we identify as a canonical ideal gas expression integrated against the Harish-Chandra character of a massive scalar field with $m^2 \ell^2 = \frac{q^2 \ell^2}{\pi}$ with its zero-mode removed  (see expression (3.10) of \cite{Anninos:2020hfj}). The quantity $\Delta$ labels the unitary irreducible representation (UIR) of the de Sitter group this scalar field falls in.

To complete our calculation of \eqref{eq:logpart}, we need to also evaluate the above expression at $q^2\ell^2=0$, corresponding to $\Delta=0$. Formally setting $\epsilon = 0$, we have: 
\begin{equation}
\label{det2}
-\frac{1}{2} \log {{{{\det}'_\epsilon} \left(-\nabla^{2}  \right)}}  = \int_0^\infty \frac{dt}{2t} \left[ \frac{1+e^{-t}}{1-e^{-t}} \frac{e^{-\frac{t}{2}}}{(1-e^{-t})} - e^{-\frac{t}{2}}\right] \left( e^{-\frac{t}{2}} + e^{\frac{t}{2}} \right) ~.
\end{equation}
Combining everything, our task is now to compute the following regulated quantity:
\begin{multline}
-\frac{1}{2} \log {\det}_{\epsilon}' \left( 1- \frac{q^2 \ell^2}{\pi} \nabla^{-2} \right)\equiv-\frac{1}{2}\log {\det}'_{{\epsilon}}\left(-{\nabla}^2+ \frac{q^2}{\pi }\right)+\frac{1}{2} \log {{{{\det}'_\epsilon} \left(-\nabla^{2}  \right)}}\\
=-2\int_{0}^\infty \frac{\dd v}{\sqrt{v^2 + \frac{{\epsilon}^2}{ \ell^{2}}}}\left( \frac{1+e^{-\sqrt{v^2 + \frac{{\epsilon}^2}{ \ell^{2}}}}}{\left(1-e^{-\sqrt{v^2 + \frac{{\epsilon}^2}{ \ell^{2}}}}\right)^2}-1\right) e^{-\frac{1}{2}\sqrt{v^2 + \frac{{\epsilon}^2}{ \ell^{2}}}}\sinh\left(\frac{v\Delta}{2}\right)\sinh\left(\frac{v\bar\Delta}{2}\right) \, .\label{detq3}
\end{multline}

We wish to extract the behavior of this formal expression in the limit as $\epsilon\rightarrow0$.

One way to do this is to take two $q^2\ell^2$ derivatives of the expression above and compute the integral by first setting $\epsilon=0$ in the integrand.  The result is finite and agrees with the formal sum over eigenvalues from the unregulated path integral. Explicitly,
\begin{align}
\partial^2_{q^2\ell^2} \left(-\frac{1}{2} \log {\det}_{\epsilon}' \left( 1- \frac{q^2 \ell^2}{\pi} \nabla^{-2} \right)\right)& = \frac{1}{2} \sum_{L=1}^\infty  \frac{2L+1}{ ({q^2\ell^2}+\pi  L (L+1))^2} \\
&= \frac{\psi ^{(1)}\left(1+\Delta\right)-\psi ^{(1)}\left(1+\bar\Delta\right)}{2 \pi^2 \left(\bar\Delta-{\Delta}\right)}~,\label{flogZ}
\end{align}
where $\psi^{(m)}(x) = \left(\frac{d}{d x}\right)^{m+1} \log \Gamma(x)$ is the Polygamma function, and we remind the reader that
\begin{equation}
\Delta \equiv \frac{1}{2}+i \sqrt{ \tfrac{q^2 \ell^2}{\pi}-\frac{1}{4}}~, \quad\quad \bar{\Delta} \equiv 1- \Delta~.
\end{equation}
We have numerically checked that the (\ref{flogZ}) indeed agrees with the second $q^2\ell^2$ derivative of (\ref{detq3}) at small but finite $\epsilon$. 

The standard structure of ultraviolet divergences in local quantum field theory dictates that the small-$\epsilon$ limit should yield divergent terms of the form $1/\epsilon^2$ or $\log\epsilon$. From the perspective of \eqref{flogZ}, these divergences can appear in $\left(-\frac{1}{2} \log {\det}_{\epsilon}' \left( 1- \frac{q^2 \ell^2}{\pi} \nabla^{-2} \right)\right)$ upon integrating twice with respect to $q^2\ell^2$. 

We fix the coefficient of the logarithmic divergence by analyzing the $1/t$ term in the small-$t$ expansion of the integrand in (\ref{detq2}), (\ref{det2}). By analyzing the integral \eqref{detq3}, systematically in the small-$\epsilon$ limit, we find: 
\begin{align}
-\frac{1}{2} \log {\det}_{\epsilon}' \left( 1- \frac{q^2 \ell^2}{\pi} \nabla^{-2} \right) = &
 i \nu \log \frac{\Gamma (1+\bar\Delta )}{\Gamma(1+\Delta)}+\left( \psi^{(-2)} (1+\Delta ) + \psi^{(-2)} (1+\bar{\Delta} ) \right)\nonumber \\
&-\left(\psi^{(-2)}(1) + \psi^{(-2)}(2) \right) - \frac{q^2 \ell^2}{\pi} \log\left( \frac{2 \ell}{e^{\gamma}\epsilon} \right)~.\label{finalZ}
\end{align}
and we remind the reader that the quantity appearing in the logarithm is none other than $\ell\Lambda_{\rm UV}^\Phi$~. Note that \eqref{finalZ} contains no leading local ultraviolet divergent terms that go as $1/\epsilon^2$, which cancels from the subtraction of \eqref{det2}, stemming from the Jacobian term $J_{\Phi,h}$. Moreover the entire expression vanishes at $q^2\ell^2=0$, as it should. 

Putting everything together, we thus have
\begin{align}
\log Z_{S^2}^{\epsilon\tilde{\epsilon}} =& \log\mathcal{N}_{\tilde{\epsilon}}^\Psi - \frac{q^2 \ell^2}{\pi} \log\left( \ell\Lambda_{\rm UV}^\Phi\right)+\log\frac{ q\ell}{2\pi}+
 i \nu \log \frac{\Gamma (1+\bar\Delta )}{\Gamma(1+\Delta)}\nonumber \\
&+\left( \psi^{(-2)} (1+\Delta ) + \psi^{(-2)} (1+\bar{\Delta} ) \right)-\left(\psi^{(-2)}(1) + \psi^{(-2)}(2) \right)~\label{eq:Ztotal}
\end{align}
with $\mathcal{N}_{\tilde{\epsilon}}^\Psi$ given in \eqref{eq:normalizationfermdet}. The perturbative expansion of $\log Z_{S^2}^{\epsilon\tilde{\epsilon}}$ can now be readily obtained by a standard Taylor expansion of \eqref{eq:Ztotal} about the free point $q^2\ell^2 = 0$. For instance at fourteenth order in the loop-expansion we have 
\begin{equation}
  \log Z_{S^2}^{\epsilon\tilde{\epsilon}} \Big\lvert_{(q^2 \ell^2)^7} = -\left( \frac{q^2 \ell^2}{\pi} \right)^7 \frac{1}{7} \left( \zeta(7) + 14 \zeta(5) + 42 \zeta(3) - 66 \right)~.
\end{equation}
Remarkably, the exact expression \eqref{eq:Ztotal} predicts that quantum loop contributions to $\log Z_{S^2}^{\epsilon\tilde{\epsilon}}$ involving three or more interaction vertices are free of ultraviolet and infrared divergences altogether.

\bibliographystyle{utphys2}
\bibliography{refs}

\end{document}